\documentclass[11pt]{article}
\usepackage{amssymb, amsmath}
\usepackage{graphicx}
\newcommand{\bm}[1]{\mbox{\boldmath $#1$}}
\begin{document}
\makeatletter
\@addtoreset{equation}{section}
\def\theequation{\thesection.\arabic{equation}}
\def\@maketitle{\newpage
 \null
 {\normalsize \tt \begin{flushright} 
  \begin{tabular}[t]{l} \@date  
  \end{tabular}
 \end{flushright}}
 \begin{center} 
 \vskip 2em
 {\LARGE \@title \par} \vskip 1.5em {\large \lineskip .5em \begin{tabular}[t]{c}\@author 
 \end{tabular}\par} 
 \end{center}
 \par
 \vskip 1.5em} 
\makeatother
\topmargin=-1cm
\oddsidemargin=1.5cm
\evensidemargin=-.0cm
\textwidth=15.5cm
\textheight=22cm
\setlength{\baselineskip}{16pt}
\title{Interior of the Horizon of BTZ Black Hole
}
\author{
Ryuichi~{\sc Nakayama}\thanks{nakayama@particle.sci.hokudai.ac.jp} and Kenji~{\sc Shiohara}\thanks{k-shiohara@particle.sci.hokudai.ac.jp}  
       \\[1cm]
{\small
    Division of Physics, Graduate School of Science,} \\
{\small
           Hokkaido University, Sapporo 060-0810, Japan}
}
\date{
EPHOU-20-010\\
}
%
%
\maketitle

\begin{abstract} 
A quantum scalar field inside the horizon of a non-rotating BTZ black hole is studied. Not only the near-horizon modes but also the normal modes deep inside the horizon are obtained.   It is shown that the matching condition for the normal modes of a scalar field at the horizon does not uniquely determine the normal-mode expansion  of a scalar field inside the horizon. By choosing a certain appropriate prescription for removing this ambiguity an integral form of a new scalar propagator for points on both sides of the horizon are obtained. 
A similar problem may arise in higher-dimensional black holes. 
\end{abstract}
\newpage
\setlength{\baselineskip}{18pt}
\section{Introduction}
\hspace*{5mm}

The interior of the black hole is not well understood. Understanding of its structure is necessary to resolve the information paradox\cite{Hawking} and the firewall problems. \cite{AMPS}\cite{AMPSS}\cite{Raju1}\cite{Raju2}\cite{Raju3} \cite{MP}\cite{H} Recently there has been progress in the study of Hawking radiation. \cite{PSSY}\cite{AHMST}

In this paper a scalar field in the Ba\~{n}ados-Teitelboim-Zanelli (BTZ) black hole\cite{BTZ} is studied and the problem of duality between the region behind the horizon and the 2d conformal field theory (CFT) on the infinite boundary is revisited. The singularity of bulk to boundary propagators of scalar field in BTZ black hole was studied in \cite{KOS} and in this paper it was shown that the region inside the horizon can be described in terms of the boundary CFT.  BTZ solution is obtained by identifying points in AdS$_3$ described by a hyperboloid embedded in a flat space with signature $(1,1,-1,-1)$: $x_0^2+x_1^2-x_2^2-x_3^2=1$.\cite{BTZ2} AdS$_3$ is classified into three types of regions and each region is covered by four separate coordinate patches. Each region has coordinates $(r,t,\phi)$ and coordinates of every pair of regions are assumed to be related by analytic continuation of variable  $t$. In \cite{Ichinose} a scalar field propagator in BTZ black hole spacetime was obtained by using a propagator in the global patch of AdS$_3$ spacetime. A bulk to boundary propagator for a point inside the horizon is obtained by analytic continuation via a shift of $t$ by $i\beta/4$, where $\beta$ is the inverse of the temperature of the boundary CFT. \cite{KOS} 

In \cite{GT} a bulk local state of a scalar field behind as well as outside the horizon was constructed.   It was also shown that a scalar field propagator in the Rindler-AdS is obtained from that in the global AdS  by analytic continuation $(t,\phi) \rightarrow (-i\phi_r, -it_r)$. Then the propagator in BTZ background was obtained by restricting the region for $\phi_r$ from $-\infty <\phi_r< \infty$ to $0 \leq \phi_r \leq 2\pi$. The propagator obtained is a universal function of a geodesic distance of the two points, which is valid even if the point of the scalar field is behind the horizon. 
Information on the matching condition for the normal modes of a scalar field at the horizon appropriate for the above propagator was not studied. 

It is also known that the metric tensor of BTZ spacetime can be obtained from that for AdS$_3$ spacetime by a coordinate transformation.\cite{Banados}  By using these coordinate transformations and the shift of $t$, it is formally possible to construct two-point functions between points inside and outside horizons from the two-point functions for both points outside horizons. The results are expressed in terms of a geodesic distance between the two points. 

These results, however, have not been obtained by a canonical method of field theory quantization in the black hole background. It is not clear whether these results can also be obtained by the canonical method, until the quantization is carried out explicitly. Importantly, the roles of the time variable $t$ and the radial variable $r$ outside the horizon are interchanged inside the horizon. 
Because the singularity of BTZ black hole, (Milne universe) $\times \mathbb{R}_t^1$, is mild, it is expected that the above problems can be studied by ordinary quantization. 

A general prescription for quantization of matter fields inside the horizon of an eternal black hole was proposed by Papadodimas and Raju\cite{Raju1}\cite{Raju2}\cite{Raju3}; we must find out normal modes of matter fields inside the horizon, which have distinct power behaviors near the horizon, and impose suitable matching conditions on the fields on both sides of the horizon. Then we use the same set of  creation and annihilation operators on both sides of the horizon.  It was shown \cite{Raju1} that this prescription works in the case of (Minkowski) Rindler space. Because there are not so many examples, where concrete calculations are possible, it is desirable to carry out quantization of matter fields inside horizon in cases of black hole spacetimes, and show whether the prescription works, or there arise any problems in the case of real black holes. Purpose of this paper is to study these problems, and try to find what kind of description in terms of CFT's on AdS boundaries the interior of BTZ black hole has according to the principle of holography\cite{Maldacena}\cite{Kabat}. This is important, because for higher dimensional black holes it is not possible to use a pure AdS spacetime  to quantize a scalar field in black hole backgrounds.  

In this paper we will use a new approach for quantization of scalar field in the spacetime of BTZ black hole. The eternal AdS-Schwartzschild balck hole is described by a tensor product state called Thermo-Field Double (TFD) introduced by Israel\cite{Israel}. 
\begin{equation}
|\Psi_{TFD} \rangle_{\beta} =\frac{1}{\sqrt{Z(\beta)}}\sum_{n} e^{-\beta E_n/2} \, |\tilde{n}\rangle_L \otimes |n\rangle_R  \label{TFD}
\end{equation}
Here $\beta$ is an inverse temperature. This thermal state is a Hartle-Hawking-like state.\cite{BH1} The spacetime is represented by Penrose diagram and it consists of four regions I, II, III and IV (Fig.1). For quantization of  a matter (scalar) field it is decomposed into a complete orthonormal set of positive-frequency modes $f_{\omega \ell m}$ on a Cauchy surface $\Sigma$ and the coefficient of the expansion is regarded as the annihilation operator $a_{\omega \ell m}$, which satisfies together with the creation operator the usual commutator algebra.\cite{BD} Usually, this is carried out in regions I and III separately, which are outside the horizon. In this case a real scalar field is represented in terms of a single set of creation and annihilation operators in region I, and in terms of another independent set in region III. Then in order to make the scalar field  smoothly connected at the horizon the state (\ref{TFD}) is constructed in the tensor-product Hilbert space.\cite{Israel}\cite{BH1}
The  normal modes in black hole background are obtained from those of AdS$_3$ spacetime by suitable coordinate transformations.\cite{Banados}  These normal modes turn out  not eigenstates of energy and momentum. We quantize a scalar field on the  constant-$t$ slice, which is obtained by combining those slices in both regions I and III, by requiring that the normal modes form a complete orthonormal set on the combined constant-$t$ slice. 
This prescription  ensures the smoothness of the scalar field at the horizon automatically. We take time $t$ in region III to flow upwards(Fig.II).  By changing basis of the normal modes to that of eigenstates of energy and momentum it is found that the creation and annihilation operators  in each regions I and III,  which are diagonalized into the eigenstates of energy and momentum, are identified as primary operators of left and right boundary CFTs. Then the vacuum state is shown to be the TFD (\ref{TFD}).  Furthermore,  in this study each normal modes are represented in terms of integral representations, and this makes analysis of explicit asymptotic forms of these normal modes easy. 

Then region II behind the horizon will be studied.  In region II the integral representations of normal modes are technically helpful. The normal modes which we will study are those present throughout behind the horizon and are distinct from the near-horizon modes. 
In this paper it is shown that normal modes for a scalar field behind the horizon are not obtained simply by a coordinate transformation of a single set of the normal modes in AdS$_3$ spacetime. In order to impose proper canonical commutation relations on the scalar field,  it is necessary to include extra linearly independent mode and also modes which correspond to the non-normalizable modes  in AdS$_3$ spacetime. It is also shown that the inner products of mode functions inside the horizon are discontinuous at the horizon. Then, it will be shown that the matching condition for the scalar field at the horizon makes the near-horizon modes on both sides of the horizon connected, but are not sufficient to uniquely determine the normal mode expansion of scalar fields deep inside the horizon.  There exist some undetermined coefficient functions in the normal mode expansion for the scalar field inside the horizon.  Unless appropriate additional conditions are imposed on the scalar theory inside the horizon, interior of the horizon will not be uniquely determined by the boundary CFT's. We will  remove this ambiguity by a boundary condition at the future and a new scalar propagator for two points inside and outside the horizon separately will be obtained. This propagator does not  agree with the results of \cite{Ichinose} and \cite{GT}. When a scalar field is quantized in backgrounds of higher-dimensional balck holes, the double analytic continuation of coordinates is unavailable. Then it will be necessary to impose boundary conditions at the horizon by the matching condition. Similar ambiguity of the interior normal modes might occur. 

This paper is organized as follows. In secs. 2 to 4 a scalar theory outside the horizon of BTZ black hole is quantized by using the method mentioned above. In sec. 5 a set of normal modes of a scalar field in region II is found, and in sec.6 it is found that there exists discontinuity in the inner products of these normal modes at the horizon. In sec.7 matching conditions for the normal modes in region II and those in regions I and III are solved. In sec. 8 this solution is analyzed. In sec.9 a prescription for determining the undetermined parameters in the mode expansion of the scalar field is proposed. New scalar field propagator for two points behind and outside the horizon, respectively, is written down.  Summary and discussions are given in sec.10. 
Some definitions and details of calculations are put in appendices A to E. 

\section{Normal Modes and Klein-Gordon Inner Product Outside Horizon}
\hspace*{5mm}
The normal mode functions of a real scalar field in  BTZ black hole background were obtained in \cite{Ichinose}, \cite{Vakkuri}. For simplicity only the BTZ black hole without angular momentum will be considered in this paper. The metric field is given by 
\begin{equation}
ds^2 = \frac{\ell^2}{r^2-a}dr^2-(r^2-a)dt^2+r^2d\varphi^2. \label{BTZ}
\end{equation}
Here $a=8GM\ell^2$ is the mass of the black hole. The horizon is located at $r=r_+=\sqrt{a}$. In the following  the AdS length $\ell$ will be set to unity. 
The classical equation of motion for a scalar field $\phi(t,r,\varphi)$ with mass $m$ is given by
\begin{equation}
(r^2-a)\partial_r^2\phi+\frac{1}{r}(3r^2-a)\partial_r\phi-\frac{1}{r^2-a}\partial_t^2\phi+\frac{1}{r^2}\partial_{\varphi}^2\phi-m^2\phi=0.  \label{modeBTZ}
\end{equation}
After separation of variables the normal mode functions of the scalar field are written as
\begin{equation}
\phi_{\omega n}(t,r,\varphi)=e^{-i\omega t+in\varphi} \, f_{\omega n}(r), \qquad (n \in \bm{ Z}). \label{wrong}
\end{equation}
It was found that those solutions, $f_{\omega n}(r) \sim r^{1+\nu}$, which satisfy the normalizable boundary condition \cite{BDHM} at the infinite boundary ($r \sim \infty$),  are given by
\begin{equation}
f_{\omega n}(r) = (u-1)^{\alpha}u^{-\alpha-\frac{\Delta}{2}}F(\alpha+\beta+\frac{\Delta}{2},\alpha-\beta+\frac{\Delta}{2},\Delta;\frac{1}{u}).  \label{KGr}
\end{equation} 
The variable $u$ is defined by $u=\frac{r^2}{a}$, and the parameters are  
\begin{equation}
\alpha=\pm i\frac{\omega}{2\sqrt{a}}, \quad \beta=\pm i\frac{n}{2\sqrt{a}}.
\end{equation}
$F(a,b;c;z)$ is a hypergeometric function.  
Here $\nu=\sqrt{1+m^2}$ and $\Delta=1+\nu$ is a scaling dimension of the boundary operator in the dual conformal field theory (CFT). Only the case of $\nu \neq$ integer will be considered in this paper. 

\begin{figure}[thb]
\hspace {4cm}     \begin{minipage}{0.5\hsize}
      \begin{center}
\includegraphics[scale=.3, clip]{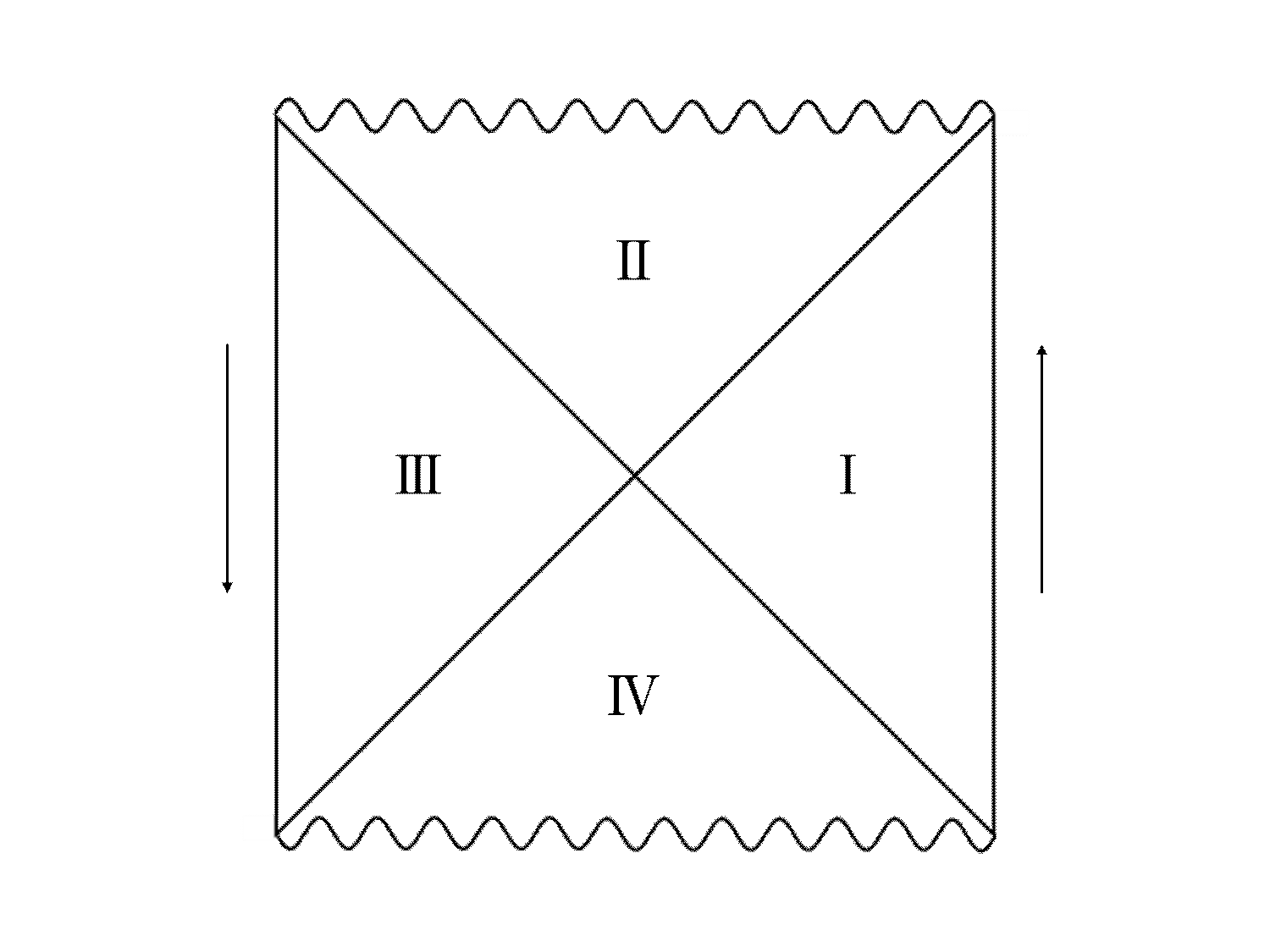}
\hspace{1.6cm} Fig. 1: Penrose diagram: on the right boundary time flows upwards and on the left boundary time flows downwards.
        \end{center}
      \end{minipage}
\end{figure}

In what follows we will use the following form of metric for the outer region of non-rotating BTZ black hole.
\begin{eqnarray}
ds^2 &=&  \frac{1}{y^2}dy^2-\frac{1}{y^2}\Big(1-\frac{a}{4}y^2\Big)^2dt^2+\frac{1}{y^2}\Big(1+\frac{a}{4}y^2\Big)^2d\varphi^2,   \label{BTZouter}
\end{eqnarray}
This is obtained from (\ref{BTZ}) by a transformation of the coordinate $r$.\cite{Banados}
\begin{equation}
r =\frac{1}{y}+\frac{a}{4}y
\end{equation}
This satisfies $r \geq \sqrt{a}$ and the horizon $r=\sqrt{a}$ corresponds to $y=\frac{2}{\sqrt{a}}$. $y=0$ and $y=\infty$ correspond to $r = \infty$, the boundary. So there are two outer regions. 
Penrose diagram of maximally extended spacetime of BTZ black hole is divided into four regions, I, II, III, IV.(Fig.1) 
The outer regions are I (right, $0<y \leq\frac{2}{\sqrt{a}}$) and III (left, $y \geq \frac{2}{\sqrt{a}}$), the inner regions are II (future) and IV (past). Normal modes of a scalar field in BTZ background are obtained from those in pure AdS$_3$ by a certain coordinate transformation. See Appendix A. The inverse temperature of this black hole is $\beta=2\pi/\sqrt{a}$. 

In the metric (\ref{BTZouter}) the angular variable $\varphi$ takes values in the range $0 \leq \varphi \leq 2\pi$. In what follows this periodic variable $\varphi$ is replaced by a line variable $x$ ($-\infty < x< \infty$) and quantization of a scalar field in the black brane will be considered. After quantization of a scalar field a scalar theory in the spherical BTZ black hole can be obtained by replacing $x$ by $\varphi+2\pi n$ and summing correlation functions over an integer $n \in \bm{Z}$. This will be done at the end of this paper. 
We will quantize a scalar field on a $t=0$ hypersurface $\Sigma$ in the space obtained by combining regions I and III. (Fig.2) The direction of $t$ in region III will be flipped compared to that in Fig.1. Region I is considered first and then region III is studied and the results will be combined. Normal modes for a scalar field in region I ($0 < y \leq \frac{2}{\sqrt{a}}$) are given by\footnote{Here $\omega$ is different from the one in (\ref{wrong}).}
\begin{multline}
\Phi^{\text{I}}_{\omega,k}(t,y,x) \equiv  \frac{4\sqrt{a}y}{4+ay^2} e^{\sqrt{a}x} J_{\nu}\Big(\sqrt{\omega^2-k^2}\frac{4\sqrt{a}y}{4+ay^2}e^{\sqrt{a}x}\Big) \\ \times
\exp \, \big[ i \frac{4-ay^2}{4+ay^2}e^{\sqrt{a}x} (k\cosh \sqrt{a}t-\omega \sinh \sqrt{a}t) \big]
\label{phiI}
\end{multline}
Here $\omega$ and $k$ are parameters which take  values in $\omega \geq 0$ and $|k| \leq \omega$.  The superscript I on $\Phi$ shows that these are modes in region I. As for dependence on $y$, $\Phi^{\text{I}}_{\omega,k}(t,y,x)  $  behaves as $y^{\Delta}$  in the limit $y \rightarrow 0$.  These modes are regular at the horizon $y=\frac{2}{\sqrt{a}}$. A mode  $\Pi=\sqrt{-g}(-g^{tt}) \partial_t\Phi_{\omega,k}^{\text{I}}$ for the momentum  conjugate to $\Phi_{\omega,k}^{\text{I}}$ is also regular at the horizon.  

Actually, both the parameters $\omega$ and $k$ are not energy and momentum. Later, the solution in the form (\ref{wrong}) will be obtained by carrying out Fourier transformations over $\zeta$ and $\mu$ in (\ref{phiI}), where $\omega$ and $k$ are related to new ones $\zeta$ and $\mu$ by
\begin{eqnarray}
\omega &=& \mu \cosh \zeta, \label{omegadef}\\
k &=& \mu \sinh \zeta. \label{kdef}
\end{eqnarray}
These new parameters take values in $-\infty <\zeta < \infty$ and $0\leq \mu <\infty$. 
In this section quantization of a scalar field is studied by using these modes. 

First, the Klein-Gordon (K-G) inner product for these modes in region I will be computed. 
The K-G inner product in regions I is defined for functions $f(t,y,x)$ and $g(t,y,x)$ as follows.
\begin{equation}
(f,g)_{\text{I}} \equiv i\int_0^{\frac{2}{\sqrt{a}}}dy \int_{-\infty}^{\infty}dx\, \frac{1}{y}\frac{4+ay^2}{4-ay^2}\Big[ f^{\ast}\partial_t g-g\partial_t f^{\ast}\Big], \label{KGinI}
\end{equation}
The integration region for $y$ is $0 \leq y \leq \frac{2}{\sqrt{a}}$.
The expression in the integrand which depends on $y$ is $\sqrt{-g}(-g^{tt})$. The subscript I for the inner product on the left hand side means that the integration is to be carried out in region I. 
This inner product does not depend on $t$. Hence this  can be computed at $t=0$. 

By setting $t=0$ and using (\ref{phiI}) the inner product $(\Phi^{\text{I}}_{\omega,k}, \Phi^{\text{I}}_{\omega',k'})_{\text{I}}$ is given by 
\begin{eqnarray}
(\Phi^{\text{I}}_{\omega,k}, \Phi^{\text{I}}_{\omega',k'})_{\text{I}}
&=& (\omega+\omega') \int_0^{\frac{2}{\sqrt{a}}} dy \int_{-\infty}^{\infty}dx \frac{16a^{\frac{3}{2}}y}{(4+ay^2)^2} e^{3\sqrt{a}x} e^{i(k'-k)\frac{4-ay^2}{4+ay^2}e^{\sqrt{a}x} }\nonumber \\
&&\times J_{\nu}\Big(\sqrt{\omega^2-k^2}\frac{4\sqrt{a}y}{4+ay^2}e^{\sqrt{a}x}\Big)J_{\nu}\Big(\sqrt{\omega'^2-k'^2}\frac{4\sqrt{a}y}{4+ay^2}e^{\sqrt{a}x}\Big). 
\end{eqnarray}
After rescaling $y \rightarrow \frac{2}{\sqrt{a}}y$, we set $\rho=\exp \sqrt{a}x$. Then we have
\begin{multline}
(\Phi^{\text{I}}_{\omega,k}, \Phi^{\text{I}}_{\omega',k'})_{\text{I}} \\
= (\omega+\omega') \int_0^2 dy \int_0^{\infty}d\rho \, \rho^2 \frac{16y}{(4+y^2)^2}  e^{i(k'-k)\frac{4-y^2}{4+y^2}\rho} J_{\nu}\Big(\mu\frac{4y}{4+y^2}\rho\Big)  
 J_{\nu}\Big(\mu'\frac{4y}{4+y^2}\rho\Big).  \label{innerproduct1.6}
\end{multline}
Here we defined 
\begin{equation}
\mu \equiv \sqrt{\omega^2-k^2}, \qquad \mu' \equiv \sqrt{\omega'^2-k'^2}.
\end{equation}
This integral appears  to be imaginary.  To check this is the case, let us change variables from $x$ to $\rho= \exp (\sqrt{a}x)$  and define new integration variables in place of $\rho$ and $y$.
\begin{equation}
z=\frac{4y\rho}{4+y^2}, \qquad w=\frac{4-y^2}{4+y^2}\rho
\end{equation}
These take values in $z, w \geq 0$ and then the above integral is evaluated as
\begin{eqnarray}
(\Phi^{\text{I}}_{\omega,k}, \Phi^{\text{I}}_{\omega',k'})_{\text{I}} 
&=& (\omega+\omega') \int_0^{\infty} dz \int_0^{\infty}dw \, z  J_{\nu}\Big(\mu z\Big)  
 J_{\nu}\Big(\mu'z\Big) \exp [-i(k-k')w]  \nonumber \\
&= &\frac{-i}{\mu}(\omega+\omega')\delta(\mu-\mu') \frac{1}{k-k'-i\epsilon} \nonumber  \\
&=& \frac{1}{\mu}(\omega+\omega')\delta(\mu-\mu')\Big( \pi \delta(k-k')-i\frac{\text{P}}{k-k'}\Big)\label{innerproductI}
\end{eqnarray}
Here P stands for a principal-value prescription  and here the following Fourier-Bessel formula is used.
\begin{equation}
\int_0^{\infty} dx \, x \, J_{\nu}(ax)J_{\nu}(bx)=\frac{1}{a}\, \delta (a-b) \qquad (a, b>0)
\end{equation}
So this inner product is a projection matrix. 

\begin{figure}[thb]
\hspace {4cm}     \begin{minipage}{0.5\hsize}
      \begin{center}
\includegraphics[scale=.3, clip]{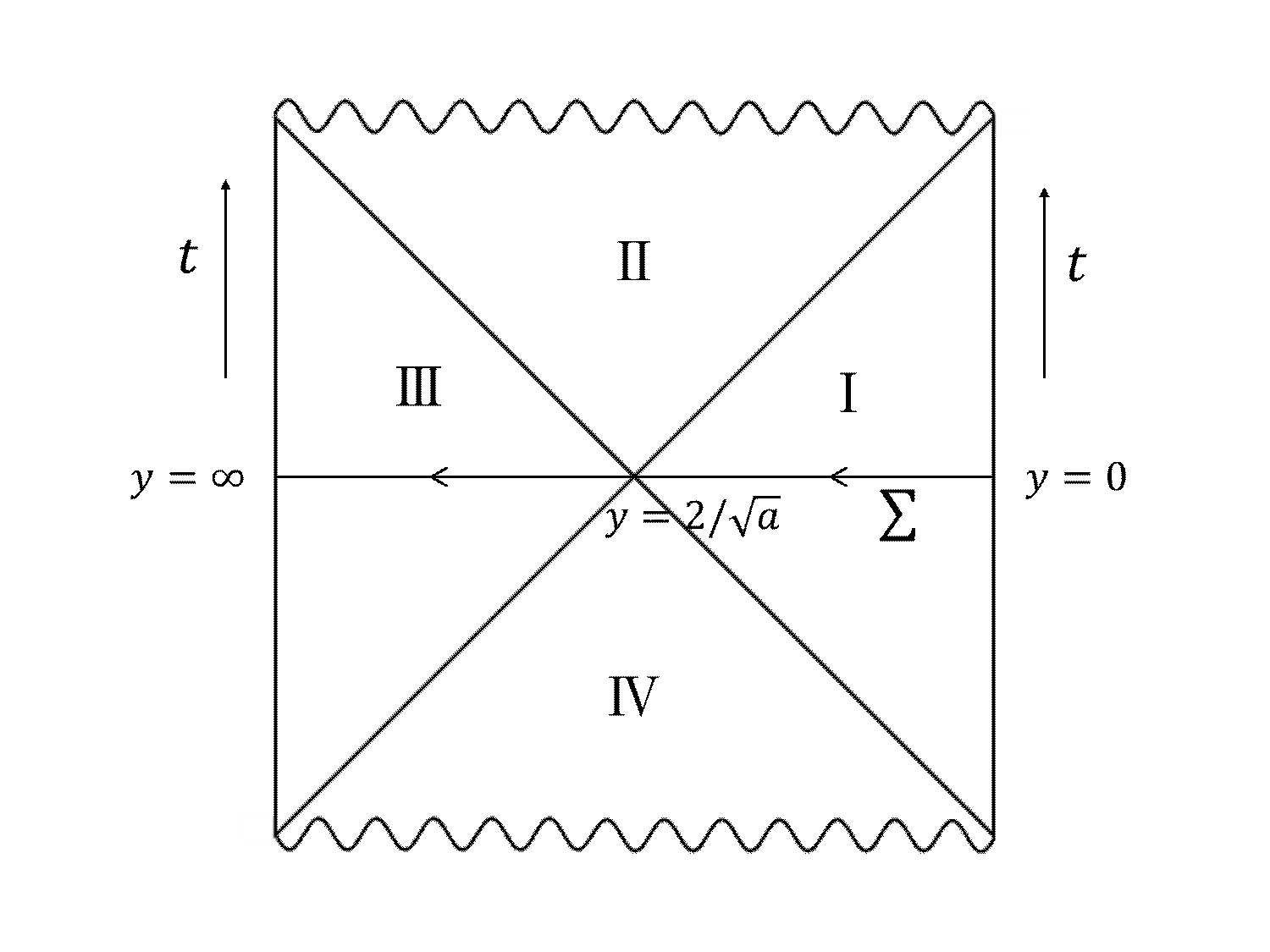}
\hspace{1.6cm} Fig.2: On both boundaries time flows upwards. $\Sigma$ is a $t=0$ Cauchy Slice. 
       \end{center}
      \end{minipage}
\end{figure}
We also need  normal modes in  region III $(2/\sqrt{a} <y$) and choose the following. 
\begin{multline}
\Phi^{\text{III}}_{\omega,k}(t,y,x) \equiv \Phi^{\text{I}}_{\omega,k}(-t,y,x)= \Big[\Phi^{\text{I}}_{\omega,-k}(t,y,x) \Big]^{\ast} \\
= \frac{4\sqrt{a}y}{4+ay^2} e^{\sqrt{a}x} J_{\nu}\Big(\sqrt{\omega^2-k^2}\frac{4\sqrt{a}y}{4+ay^2}e^{\sqrt{a}x}\Big) \\ 
\exp \big[ i \frac{4-ay^2}{4+ay^2}e^{\sqrt{a}x} (k\cosh \sqrt{a}t+\omega \sinh \sqrt{a}t)\big] 
\label{phiIII}
\end{multline}
We flip the direction of time $t$ in region III compared to that in Penrose diagram of Fig.1 and  $t$ is assumed to flow upwards. (Fig.2)
We will quantize the scalar field on the $t=0$ Cauchy surface $\Sigma$. As for dependence on $y$, $\Phi^{\text{III}}_{\omega,k}(t,y,x) $ behaves as  $y^{-\Delta}$ in the limit $y \rightarrow \infty$. 
It can be shown that  the following two sets of normal modes in the full outer region, ($0 < y <\infty$),   form a complete set of orthonormal functions, although the two regions are causally disconnected. 
\begin{equation}
\Phi_{\omega,k}(t,y,x) \equiv  \left\{ \begin{array} {cc}
\Phi^{\text{I}}_{\omega,k}(t,y,x) & (0<y<\frac{2}{\sqrt{a}}), \\
\Phi^{\text{III}}_{\omega,k}(t,y,x) & (\frac{2}{\sqrt{a}}<y) \end{array}\right.
\end{equation}
At the horizon and $t=0$ these modes are smoothly connected. 
In  region III we have an inner product, 
\begin{equation}
(f,g)_{\text{III}} \equiv i\int_{\frac{2}{\sqrt{a}}}^
{\infty}dy \int_{-\infty}^{\infty} dx \, \frac{1}{y}\frac{4+ay^2}{ay^2-4}\Big[ f^{\ast}\partial_t g-g\partial_t f^{\ast} \Big]
\end{equation}
Here the sign of the $y$-dependent factor is changed compared to (\ref{KGinI}), because integration region for $y$ is  $\frac{2}{\sqrt{a}} \leq y$.
The inner product of $\Phi^{\text{III}}_{\omega,k}$ after rescaling $y \rightarrow \frac{2}{\sqrt{a}}y $ and change of variable $\rho=exp \sqrt{a}x$ is given by  
\begin{multline}
(\Phi^{\text{III}}_{\omega,k}, \Phi^{\text{III}}_{\omega',k'})_{\text{III}} \\
= (\omega+\omega') \int_2^{\infty} dy \int_0^{\infty}d\rho \, \rho^2 \frac{16y}{(4+y^2)^2}  e^{i(k'-k)\frac{4-y^2}{4+y^2}\rho} J_{\nu}\Big(\mu\frac{4y}{4+y^2}\rho\Big)   J_{\nu}\Big(\mu'\frac{4y}{4+y^2}\rho\Big).
\end{multline}
This coincides with a complex conjugate of the inner product (\ref{innerproduct1.6}). The sum of these two is a delta function.
\begin{equation}
(\Phi_{\omega,k}, \Phi_{\omega',k'})\equiv (\Phi^{\text{I}}_{\omega,k}, \Phi^{\text{I}}_{\omega',k'})_{\text{I}} + (\Phi^{\text{III}}_{\omega,k}, \Phi^{\text{III}}_{\omega',k'})_{\text{III}} =4\pi \delta(\omega-\omega')\delta(k-k')  \label{innerproduct1}
\end{equation}
As for the other inner products, it can be shown that 
\begin{equation}
(\Phi^{\text{I}\ast}_{\omega,k}, \Phi^{\text{I}}_{\omega',k'})_{\text{I}} = -
(\Phi^{\text{III}\ast}_{\omega,k}, \Phi^{\text{III}}_{\omega',k'})_{\text{III}}
=\frac{-i}{\mu} \, \delta(\mu-\mu') \, \frac{\omega-\omega'}{k+k'-i\epsilon}
\end{equation}
and $(\Phi^{\ast}_{\omega,k}, \Phi_{\omega',k'})=0$. 

The scalar field $\Phi(t,y,x)$ is expanded into modes (\ref{phiI}) and (\ref{phiIII}). In region I, we expand $\Phi$ by using $\Phi^{\text{I}}_{\omega,k}$ 
\begin{equation}
\Phi(t,y,x) =\int_0^{\infty} \frac{d\omega}{2\sqrt{\pi}} \int_{-\omega}^{\omega} dk \Big( a(\omega,k) \Phi^{\text{I}}_{\omega,k}(t,y,x)+
 a^{\dagger}(\omega,k) \Phi^{\text{I}\ast}_{\omega,k}(t,y,x) \Big),  \label{PhiI}
\end{equation}
while in region III we have, 
\begin{equation}
\Phi(t,y,x) =\int_0^{\infty} \frac{d\omega}{2\sqrt{\pi}} \int_{-\omega}^{\omega} dk \Big( a(\omega,k) \Phi^{\text{III}}_{\omega,k}(t,y,x)+
 a^{\dagger}(\omega,k) \Phi^{\text{III}\ast}_{\omega,k}(t,y,x) \Big). \label{PhiIII}
\end{equation}
The scalar field is represented in terms of the {\it single set of operators}, $a(\omega,k)$, $a^{\dagger}(\omega,k)$ in both regions I and III.
By using the K-G inner products, the creation and annihilation operators are expressed as
\begin{eqnarray}
a(\omega,k) &=& \frac{1}{2\sqrt{\pi}} \Big[ (\Phi^{\text{I}}_{\omega,k},\Phi)_{\text{I}}+ (\Phi^{\text{III}}_{\omega,k},\Phi)_{\text{III}} \Big], \label{a}\\
a^{\dagger}(\omega,k) &=& \frac{-1}{2\sqrt{\pi}} \Big[ (\Phi^{\text{I}\ast}_{\omega,k},\Phi)_{\text{I}}+ (\Phi^{\text{III}\ast}_{\omega,k},\Phi)_{\text{III}} \Big]  \label{adagger}
\end{eqnarray}
The commutation relations of these operators
\begin{eqnarray}
[a(\omega,k), a^{\dagger}(\omega',k')] &=& \delta(\omega-\omega') \delta(k-k'), \nonumber \\
\, [a(\omega,k), a(\omega',k') ] 
&=& [ a^{\dagger}(\omega,k), a^{\dagger}(\omega',k') ]=0  \label{CR}
\end{eqnarray}
 are obtained by imposing the canonical commutation relations (CCR's):
\begin{equation}
 [\Phi(t,y,\varphi),\Pi(t,y',\varphi')]=i\delta(y-y')\delta(x-x') \label{CCR1}
\end{equation}
when $(t,y,x)$ and $(t,y',x') $ are both in region I or III, and 
\begin{equation}
 [\Phi(t,y,x), \Pi(t,y',x')]=0, \label{CCR2}
\end{equation}
when the two points are separated by the horizon. Furthermore, 
\begin{equation}
  [\Phi(t,y,x), \Phi(t,y',x')]=[\Pi(t,y,x), \Pi(t,y',x')]=0 \label{CCR3}
\end{equation}
must hold for any separation of the two points. 
Here $\Pi=\sqrt{-g}(-g^{tt})\partial_t \Phi$ is a canonical momentum field. 
We also checked that these CCR's (\ref{CCR1})-(\ref{CCR3}) are satisfied for $0 < y,y'<\infty$ by using the mode expansions (\ref{PhiI})-(\ref{PhiIII}) and commutation relations (\ref{a})-(\ref{adagger}). 

The above annihilation operator defines a vacuum $|\Psi_{TFD}\rangle_{\beta}$. 
\begin{equation}
a(\omega,k)|\Psi_{TFD}\rangle_{\beta} =0 \qquad (\omega \geq 0, \ |k| \leq \omega) \label{vaccond}
\end{equation}
Here $\beta=2\pi/\sqrt{a}$ is an inverse temperature of the black hole. 
As will be shown in sec.4, this is a TFD state. 
This vacuum is invariant under $t$ and $x$ translations. Under these transformations the normal mode transforms as 
\begin{eqnarray}
\Phi^{\text{I}}_{\omega,k}(t+\epsilon,y,x) &=& \Phi^{\text{I}}_{\omega', \, k'}(t,y,x), \\
\Phi^{\text{I}}_{\omega,k}(t,y,x+\epsilon) &=&e^{\sqrt{a}\epsilon}\, \Phi^{\text{I}}_{\omega'',k''}(t,y,x), 
\end{eqnarray}
where $\omega'=\omega \cosh \sqrt{a}\epsilon+k\sinh\sqrt{a}\epsilon$, $k'=k \cosh\sqrt{a}\epsilon-\omega \sinh\sqrt{a}\epsilon$ and $\omega''= \omega \, e^{\sqrt{a}\epsilon}$, $k''=k \, e^{\sqrt{a}\epsilon}$.
Same transformations are also valid for $\Phi^{\text{III}}_{\omega,k}$. Therefore,  $a(\omega,k)$ and $a^{\dagger}(\omega,k)$ must be linear representations of the two translations, and the vacuum in  (\ref{vaccond}) respects  translation invariances of the vacuum.

\section{Change of Basis of Normal Modes}
\hspace*{5mm}
The parameters $\omega$ and $k$ which is used to label the normal modes in the previous section  do not represent the energy and momentum of the scalar particle. 
In this section the basis of the normal modes will be changed to eigenstates of energy and momentum. 
We will denote $\omega$ and $k$ as 
\begin{equation}
\omega =\mu \cosh \zeta, \qquad k=\mu \sinh \zeta, \qquad (\mu \geq 0, \ -\infty < \zeta < \infty)
\end{equation}
Because $k \cosh \sqrt{a}t-\omega\sinh \sqrt{a}t=\mu \sinh (\zeta-\sqrt{a}t)$,  it is easy to show that $\zeta$ in  (\ref{phiI}) and (\ref{phiIII}) plays the same roles as  $\sqrt{a} t$. So we carry out  the following Fourier transformations. 
\begin{eqnarray}
\Phi_{E,p}^{\text{I}} (t,y,x) &=& \int_{-\infty}^{\infty}d\zeta\int_0^{\infty}d\mu \, 
e^{-\frac{iE}{\sqrt{a}}\zeta-\frac{ip}{\sqrt{a}}  \ln \mu   } \ \Phi^{\text{I}}_{\omega,k}(t,y,x),  \label{Fourier1}\\
\Phi_{E,p}^{\text{III}}  (t,y,x) &=& \int_{-\infty}^{\infty}d\zeta\int_0^{\infty}d\mu \, 
e^{-\frac{iE}{\sqrt{a}}\zeta-\frac{ip}{\sqrt{a}}  \ln \mu   } \ \Phi^{\text{III}}_{\omega,k}(t,y,x). \label{Fourier2}
\end{eqnarray}

The new mode $\Phi_{E,p}^{\text{I}} (t,y,x)  $ is periodic function of $t$ with an imaginary period $i \frac{2\pi}{\sqrt{a}} \equiv i\beta$, as   (\ref{phiI}) is.  Now let us carry out the following shift of integration variables, $\zeta$ and $\ln \mu$. 
\begin{eqnarray}
\zeta &\rightarrow& \zeta +\sqrt{a} t, \\
\ln \mu &\rightarrow& \ln \mu -\sqrt{a}x
\end{eqnarray}
We have
\begin{equation}
\Phi_{E,p}^{\text{I}} (t,y,x) = e^{-iEt+ipx} \, \int_{-\infty}^{\infty}d\zeta\int_0^{\infty}d\mu \, e^{-\frac{iE}{\sqrt{a}}\zeta-\frac{ip}{\sqrt{a}}  \ln \mu   } \ \frac{4\sqrt{a}y}{4+ay^2}
J_{\nu}\Big(\mu \frac{4\sqrt{a}y}{4+ay^2}\Big) \, e^{i\mu\frac{4-ay^2}{4+ay^2} \, \sinh \zeta}  \label{phiIEp}
\end{equation}
Now energy $E$ and  momentum $p$ are diagonalized. This normal mode will coincide with one of (\ref{wrong}).  In this representation, the imaginary periodicity of $t$ is lost. This means that the move of the contour of  $\zeta$ in the imaginary direction is not allowed. 
At the horizon $y=\frac{2}{\sqrt{a}}$ (\ref{phiIEp})  is proportional to 
$\delta(E)$ and vanishes for $E \neq 0$. In  the AdS$_3$/CFT$_2$ correspondence  and in the leading order of $1/N$ expansion, the eigenvalues  $E$ and $p$ for a scalar field are given by $E=\Delta+2n+|m|$ and $p=m$, such that $n=0,1, \ldots$ and $m=0,\pm1, \pm 2,\ldots$. However, in the large $N$ limit, the energy eigenvalue becomes continuous.\cite{Raju1} In this paper continuous spectrum of energy and momentum eigenvalue will be adopted. 

The other normal mode $\Phi_{E,p}^{\text{III}}$ is similarly given by
\begin{equation}
\Phi_{E,p}^{\text{III}} (t,y,x) = e^{iEt+ipx} \, \int_{-\infty}^{\infty}d\zeta\int_0^{\infty}d\mu \, e^{-\frac{iE}{\sqrt{a}}\zeta-\frac{ip}{\sqrt{a}}  \ln \mu   } \ 
\frac{4\sqrt{a}y}{4+ay^2}J_{\nu}\Big(\mu \frac{4\sqrt{a}y}{4+ay^2}\Big) \, e^{i\mu\frac{4-ay^2}{4+ay^2} \, \sinh \zeta}  \label{phiIIIEp}
\end{equation}
Time $t$ in region III is defined to flow upwards as opposed to the usual choice for the Penrose diagram. 
The coefficient of $iEt$ in the first exponent is flipped w.r.t. that in (\ref{phiIEp}). 

Next the K-G inner product will be worked out.
This is done by using (\ref{innerproduct1}), (\ref{Fourier1}) and (\ref{Fourier2}). 
We have 
\begin{eqnarray}
&& (\Phi^{\text{I}}_{E,p}, \Phi^{\text{I}}_{E',p'})_{\text{I}} + 
 (\Phi^{\text{III}}_{E,p}, \Phi^{\text{III}}_{E',p'})_{\text{III}} 
\nonumber \\
&=&  \int_{-\infty}^{\infty}d\zeta \int_0^{\infty}d\mu  \, 
e^{\frac{iE}{\sqrt{a}}\zeta+\frac{ip}{\sqrt{a}}  \ln \mu   } \ \int_{-\infty}^{\infty}d\zeta' \int_0^{\infty}d\mu'  \, 
e^{-\frac{iE'}{\sqrt{a}}\zeta-\frac{ip'}{\sqrt{a}}  \ln \mu   } \ \frac{4\pi}{\mu} \delta(\zeta-\zeta')\delta(\mu-\mu') \nonumber \\
&=& 4\pi a\delta(E-E')\delta(p-p').   \label{3.8}
\end{eqnarray}  
Here a formula $\delta(\omega-\omega')\delta(k-k')=\mu^{-1}\delta(\zeta-\zeta')\delta(\mu-\mu')$ is used. 

The scalar field is expanded into the above modes (\ref{phiIEp}) and (\ref{phiIIIEp}) as follows.  In region I we have 
\begin{multline}
\Phi^{\text{I}}(t,y,x) = 1/(4\pi\sqrt{\pi a}) \int_{-\infty}^{\infty}dp \int_{0}^{\infty}dE \Big( b_+(E,p) \Phi^{\text{I}}_{E,p}(t,y,x)+
b_+^{\dagger}(E,p) \Phi^{\text{I}\ast}_{E,p}(t,y,x)  \\
  +  b_-(E,p) \Phi^{\text{I}}_{-E,-p}(t,y,x)+
b_-^{\dagger}(E,p) \Phi^{\text{I}\ast}_{-E,-p}(t,y,x)
\Big)   \label{PhibI}
\end{multline}
Here $b_{\pm}(E,p)$ are an annihilation operator for positive (negative) frequency normal modes. 
A notable point is that in the second line the negative-frequency mode $\Phi^{\text{I}}_{-E,-p} $ is associated with the annihilation operator $b_-(E,p) $.   $b_-^{\dagger}$ creates a 'hole', while $b_+^{\dagger}$ creates a `particle'.\cite{Takahashi}\cite{Umezawa} 
In sec.4 it will be shown that $\Phi^{\text{I}}_{E,p}$ and $\Phi^{\text{I}\ast}_{-E,-p}$ are linearly dependent. 

Similarly  in region III, we have 
\begin{multline}
\Phi^{\text{III}}(t,y,x) = 1/(4\pi\sqrt{\pi a}) \int_{-\infty}^{\infty}dp \int_{0}^{\infty}dE \Big( b_+(E,p) \Phi^{\text{III}}_{E,p}(t,y,x)+
b_+^{\dagger}(E,p) \Phi^{\text{III}\ast}_{E,p}(t,y,x)  \\
  +  b_-(E,p) \Phi^{\text{III}}_{-E,-p}(t,y,x)+
b_-^{\dagger}(E,p) \Phi^{\text{III}\ast}_{-E,-p}(t,y,x)
\Big)  \label{PhibIII}
\end{multline}
Therefore in region III operator $b_+$ is associated with the negative-frequency modes, and $b_-$ the positive-frequency modes. Therefore the scalar field contains negative-frequency operators as well as positive-frequency ones.  $\Phi^{\text{III}}_{E,p}$ and $\Phi^{\text{III}\ast}_{-E,-p}$ will be also found to be linearly dependent. 

$b_{+,-}$ annihilate the vacuum $|\Psi_{TFD}\rangle_{\beta}$. 
The relation between $a(\omega, k)$ and $b_{\pm}(E,p)$ is given by 
\begin{equation}
a(\omega, k)=\frac{1}{2\pi \sqrt{a}\mu} \int_0^{\infty} dE\int_{-\infty}^{\infty}dp \Big[ 
e^{-\frac{i}{\sqrt{a}}E\zeta-\frac{i}{\sqrt{a}}p\ln \mu}b_+(E,p)+e^{\frac{i}{\sqrt{a}}E\zeta+\frac{i}{\sqrt{a}}p\ln \mu}b_-(E,p) \Big]
\end{equation}
Hermitian conjugate of this equation gives a relation for $a^{\dagger}(\omega, k)$. 
The commutation relations of $b_{\pm}$ and $b_{\pm}^{\dagger}$ are given by 
\begin{equation}
 [b_+(E,p), b_+^{\dagger}(E',p')]= [b_-(E,p), b_-^{\dagger}(E',p') ]=\delta(E-E')\delta(p-p').               \label{bbCR}
\end{equation}
Other commutators vanish.  

\section{Boundary limit of the Scalar Field and Bulk Reconstruction}
\hspace*{5mm}
In this section the boundary limit of $\Phi^{\text{I}, \text{III}}$ will be considered, and the CFT operators on the two boundaries will be identified. Some detailed equations are summarized in Appendix B 

First, the normal modes $\Phi^{\text{I}}_{E,p}$ (\ref{phiIEp}) has the following  $y \rightarrow 0$ limit in accord with the BDHM dictionary.\cite{BDHM}.
\begin{equation}
\Phi^{\text{I}}_{E,p} \rightarrow y^{\Delta} \, e^{-iEt+ipx} \, g(E,p),  \label{asymp1}
\end{equation}
where $g(E,p)$ is a function defined by
\begin{equation}
g(E,p)=2^{-\nu} \, a^{\Delta/2} (\Gamma(\Delta))^{-1}\int_{-\infty}^{\infty}d\zeta \int_0^{\infty} d\mu \, \mu^{\nu} \exp \{
-i\frac{E}{\sqrt{a}}\zeta-i\frac{p}{\sqrt{a}}\ln \mu+i\mu \sinh \zeta\}  \label{gEp}
\end{equation}
Then the scalar field in region I has the following $y \rightarrow 0$ limit.
\begin{multline}
\Phi^{\text{I}}(t,y,x) \rightarrow  1/(4\pi\sqrt{\pi a})y^{\Delta}\int_0^{\infty} dE \int_{-\infty}^{\infty}dp \\
\big[ |g(E,p)|^2-|g(-E,-p)|^2\big]^{1/2}\big[ e^{-iEt+ipx} 2^{-\frac{i}{\sqrt{a}}p}\, c_R(E,p)+e^{iEt-ipx} 2^{\frac{i}{\sqrt{a}}p}\, c^{\dagger}_R(E,p)\big] \\
\equiv y^{\Delta} O_R(t,x)  \label{cO}
\end{multline}
where $O_R(t,x)$ is a CFT operator on the boundary and the operator $c_R(E,p)$ is defined by
\begin{equation}
c_R(E,p)=2^{\frac{i}{\sqrt{a}}p}\,  \big[b_+(E,p)g(E,p)+b_-^{\dagger}(E,p) g^{\ast}(-E,-p)\big] 
\big[ |g(E,p)|^2-|g(-E,-p)|^2\big]^{-1/2}.     \label{cR}
\end{equation}
$c_R^{\dagger}$ is its hermitian conjugate. These operators are boundary CFT operators on the right boundary, and satisfy the commutation relations. 
\begin{equation}
[c_R(E,p), c_R^{\dagger}(E',p')]=\, \delta(E-E')\delta(p-p')  \label{FourierR}
\end{equation}
$c_R$ is determined by boundary CFT operator $O_R(t,x)$ by Fourier transformation.\cite{Raju1} 

Similarly, the boundary limit $y \rightarrow \infty$ of the normal modes (\ref{phiIEp}) in region III is given by 
\begin{equation}
\Phi^{\text{III}}_{E,p} \rightarrow (4/a)^{\Delta} y^{-\Delta} \, e^{iEt+ipx} \, \tilde{g}(E,p),
\end{equation}
where $\tilde{g}(E,p)$ is a function defined by
\begin{equation}
\tilde{g}(E,p)=2^{-\nu} \, a^{\Delta/2} (\Gamma(\Delta))^{-1}\int_{-\infty}^{\infty}d\zeta \int_0^{\infty} d\mu \, \mu^{\nu} \exp \{i\frac{E}{\sqrt{a}}\zeta-i\frac{p}{\sqrt{a}}\ln \mu+i\mu \sinh \zeta\}  \label{tildegEp}
\end{equation}
Then the scalar field in region III has the following $y \rightarrow \infty$ limit.
\begin{multline}
\Phi^{\text{III}}(t,y,x) \rightarrow  (4/a)^{\Delta}   y^{-\Delta} \, \frac{1}{4\pi\sqrt{\pi a}} \int_0^{\infty} dE \int_{-\infty}^{\infty}dp \\
\big[|\tilde{g}(-E,-p)|^2-   |\tilde{g}(E,p)|^2\big]^{1/2} \big[ e^{iEt+ipx} 2^{\frac{i}{\sqrt{a}}p}\, c_L(E,p)+e^{-iEt-ipx} 2^{-\frac{i}{\sqrt{a}}p}\, c^{\dagger}_L(E,p)\big],
\end{multline}
where the operator $c_L(E,p)$ is defined by
\begin{equation}
c_L(E,p)=2^{-\frac{i}{\sqrt{a}}p}\, \big[ b_+^{\dagger}(E,p)\tilde{g}^{\ast}(E,p)+b_-(E,p) \tilde{g}(-E,-p) \big] \big[|\tilde{g}(-E,-p)|^2-   |\tilde{g}(E,p)|^2\big]^{-1/2}.    \label{cL}
\end{equation}
$c_L$ and $c_L^{\dagger}$ satisfy 
\begin{equation}
[c_L(E,p),c_L^{\dagger}(E',p')]=  \delta(E-E')\delta(p-p'). \label{FourierL}
\end{equation}
Moreover, because regions I and III are causally disconnected, the scalar fields in both regions commute.  Therefore $c_R$ and $c_R^{\dagger}$ commute with $c_L$, and $c_L^{\dagger}$.  $c_L$ and $c_L^{\dagger}$ are CFT operators on the left boundary.

When (\ref{cR}) and (\ref{cL}) are simplified by using the results in Appendix B, it is found that these relations are Bogoliubov transformations.  
\begin{eqnarray}
c_R(E,p) &=&  \frac{1}{\sqrt{1-e^{-\beta E}}}\Big( b_+(E,p)+e^{-(\beta/2) E}b_-^{\dagger}(E,p)\Big), \label{cb1}\\
c_L(E,p)  &=&  \frac{1}{\sqrt{1-e^{-\beta E}}}\Big( b_-(E,p)+e^{-(\beta/2) E}b_+^{\dagger}(E,p)\Big), \label{cb2}
\end{eqnarray}
\begin{equation}
c_R^{\dagger}c_R-c_L^{\dagger}c_L=b_+^{\dagger}b_+-b_-^{\dagger}b_-
\end{equation}
The thermal expectation values of the number operators are the Bose-Einstein distribution.
\begin{eqnarray}
{}_{\beta}\langle \Psi_{TFD}| c_R^{\dagger}(E,p)c_R(E',p')|\Psi_{TFD}\rangle_{\beta}&=&{}_{\beta}\langle \Psi_{TFD}| c_L^{\dagger}(E,p)c_L(E',p')|\Psi_{TFD}\rangle_{\beta} \nonumber \\
&& =\frac{1}{e^{\beta E}-1}\delta(E-E')\delta(p-p'), \nonumber \\
{}_{\beta}\langle \Psi_{TFD}| c_R(E,p)c_R^{\dagger}(E',p')|\Psi_{TFD}\rangle_{\beta}&=&{}_{\beta}\langle \Psi_{TFD}| c_L(E,p)c_L^{\dagger}(E',p')|\Psi_{TFD}\rangle_{\beta}\nonumber \\
&&=\frac{1}{1-e^{-\beta E}}\delta(E-E')\delta(p-p').  \label{VEV}
\end{eqnarray}
There are also non-vanishing thermal average's: ${}_{\beta}\langle \Psi_{TFD}| c_R(E,p)c_L(E',p')|\Psi_{TFD}\rangle_{\beta}=
{}_{\beta}\langle \Psi_{TFD}| c_L(E,p)c_R(E',p')|\Psi_{TFD}\rangle_{\beta}=e^{-\beta E/2}/(1-e^{-\beta E})$, and similar averages for $c_L^{\dagger}$ and $c_R^{\dagger}$. 
Let us define a ground state of CFT's,  $|0\rangle=|0\rangle_L \otimes |0\rangle_R$ as the state annihilated by $c_{R,L}$.
\begin{equation}
c_R|0\rangle_R=c_L|0\rangle_L=0
\end{equation}
Then by solving (\ref{cb1}), (\ref{cb2}) in favor of $b_{+,-}$ it can be shown that the state $|\Psi_{TFD}\rangle_{\beta}$ annihilated by $b_{+,-}$  is a Hartle-Hawking state, or TFD\cite{Israel}.  
\begin{equation}
|\Psi_{TFD}\rangle_{\beta} =\frac{1}{\sqrt{Z(\beta)}} \, \exp \Big[ \int_0^{\infty}dE\int_{-\infty}^{\infty}dp \,  e^{-(\beta/2)E} \, c_L^{\dagger}(E,p)c_R^{\dagger}(E,p)  \Big]\, |0\rangle_L \otimes |0\rangle_R
\label{HH}
\end{equation}
This is an entangled state. 
The Thermo-Field Hamiltonian is given by
\begin{multline}
H_{\text{TF}}=\int_0^{\infty}dE \, \int_{-\infty}^{\infty}dp \, E \, \big(b_+^{\dagger}(E,p) b_+(E,p)- b_-^{\dagger}(E,p) b_-(E,p)    \big)\\
=\int_0^{\infty}dE \, \int_{-\infty}^{\infty}dp \, E \, \big( c_R^{\dagger}(E,p) c_R(E,p)    -c_L^{\dagger}(E,p) c_L(E,p)\big) \equiv H_R-H_L.
\end{multline} 
So by carrying out quantization of a scalar field in regions I and III in a single Hilbert space we obtained a Hartle-Hawking-like state as a `vacuum' which is annihilated by the annihilation operators $b_{+,-}(E,p)$ for the scalar field. 

By using the asymptotics (\ref{asymp1}) and (\ref{gfunc}) it can be shown that as $ y \rightarrow 0$ the following equation holds. 
\begin{equation}
 \Phi^{\text{I}}_{-E,-p}-e^{-\beta E/2}\Phi^{\text{I}\ast}_{E,p}=O(y^{\Delta +1}) \label{asy}
\end{equation}
Because $\Phi^{\text{I}}_{E,p}$ is a solution to the K-G equation and the allowed asymptotic leading powers of $y$ are $1 \pm \nu$,  the left hand side vanishes identically:
\begin{equation}
\Phi^{\text{I}}_{-E,-p}-e^{-\beta E/2}\Phi^{\text{I}\ast}_{E,p}=0.  \label{4.21}
\end{equation}
Hence after some algebra the following eq is obtained.
\begin{equation}
\Phi^{\text{I}}(t,y,x)= 1/(4\pi\sqrt{\pi a}) \int_{-\infty}^{\infty}dp\int_0^{\infty}dE \sqrt{1-e^{-\beta E}} \big [ c_R(E,p)\Phi^{\text{I}}_{E,p}(t,y,x)+h.c.\big]  \label{PhiR4}
\end{equation}
This can be also expressed in terms of $O_R$ by using (\ref{cO}). Here it will not be attempted to rewrite (\ref{PhiR4})  in terms of an integral  as in \cite{Kabat}.
Time evolution of $\Phi^{\text{I}}$ is generated  by the Hamiltonian $H_R$.

Similarly in region III we obtain
\begin{equation}
\Phi^{\text{III}}_{-E,-p}-e^{\beta E/2}\Phi^{\text{III}\ast}_{E,p}=0  \label{4.23}
\end{equation}
and
\begin{equation}
\Phi^{\text{III}}(t,y,x)= 1/(4\pi\sqrt{\pi a}) \int_{-\infty}^{\infty}dp\int_0^{\infty}dE \sqrt{1-e^{-\beta E}} \big [c_L(E,p)\Phi^{\text{III}}_{-E,-p}(t,y,x)+h.c.\big] \label{PhiL4}
\end{equation}
Operator $c_R^{(\dagger)}$  exists only in region I
and $c_L^{(\dagger)}$  only in region III. By changing the basis of the normal modes to the eigenstates of energy the operators in region I and those in region III are decoupled. 
(\ref{PhiR4}) and (\ref{PhiL4}) are the bulk reconstruction of a scalar field in BTZ black hole. 

\section{Normal Modes of a Real Scalar Field behind the Horizon}
\hspace*{5mm}
In this section solutions to the K-G equation behind the horizon will be considered and the normal modes will be identified. Scalar field in the interior of the horizon will be quantized by choosing the operators to be in the same Hilbert space as those in regions I and III. Later the theories outside and inside the horizon will be connected by the matching condition\cite{Raju1}. For this purpose the normal modes of scalar filed behind the horizon must be identified. This is carried out in this section. In this paper only region II will be considered. Region IV can be similarly studied.  It is shown that the choice of the normal modes inside the horizon is complicated. It will be shown that it is necessary to take into account four types of normal modes (\ref{fourmodes}) for quantization.  

The metric inside the horizon can be obtained from (\ref{BTZ}) by a coordinate transformation.
\begin{equation}
\eta=\frac{2}{r}\Big(1\pm \frac{1}{\sqrt{a}}\sqrt{a-r^2}\Big)  \label{etapm}
\end{equation}
Then the  metric tensor behind the horizon is written as
\begin{equation}
ds^2= -\frac{16a}{(a\eta^2+4)^2}d\eta^2+\frac{a(a\eta^2-4)^2}{(a\eta^2+4)^2}dt^2+\frac{16a^2\eta^2}{(a\eta^2+4)^2}dx^2.
\end{equation}
\begin{figure}[thb]
\hspace {4cm}     \begin{minipage}{.5\hsize}
      \begin{center}
\includegraphics[scale=.3, clip]{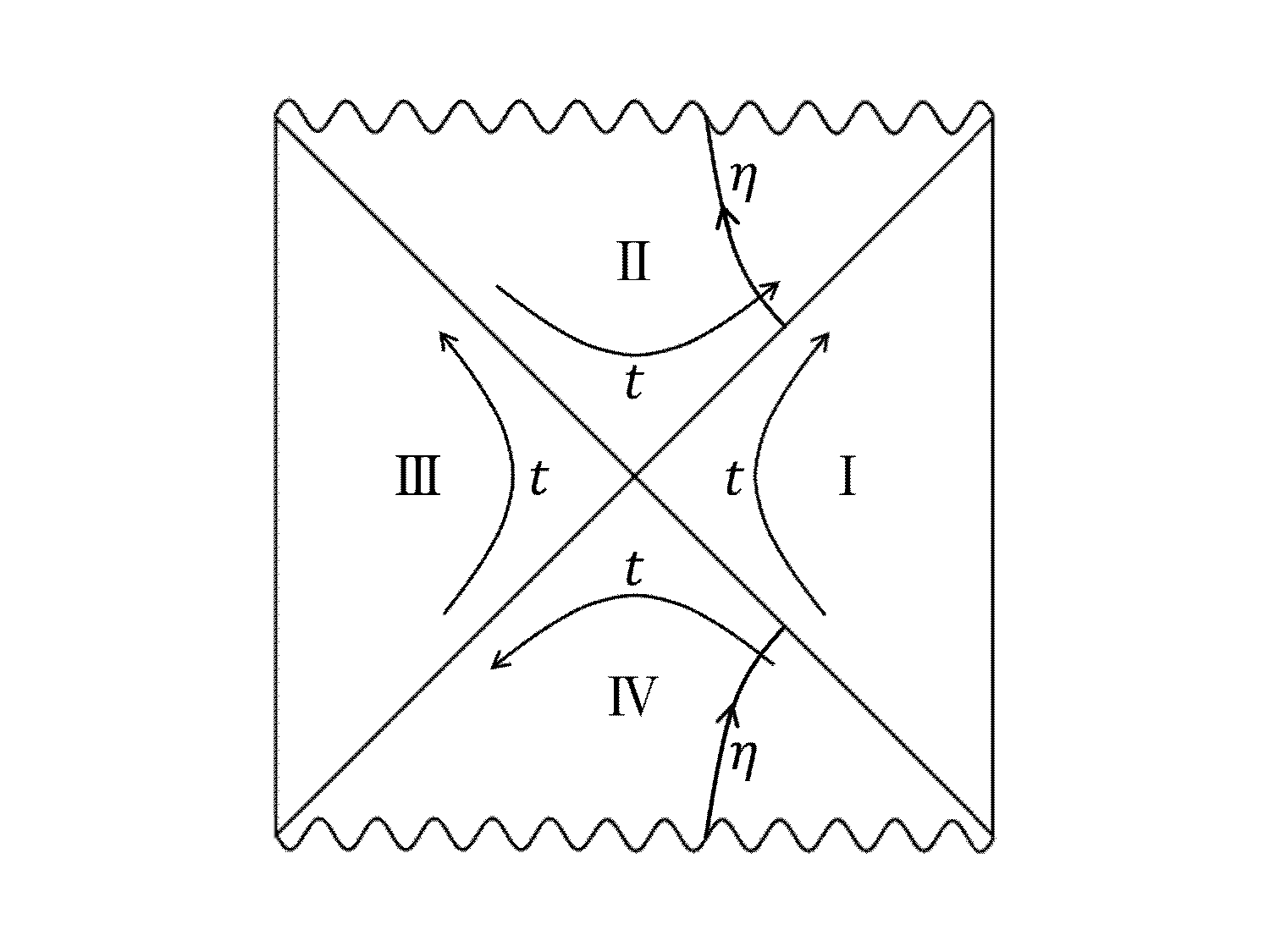}
\hspace{1.6cm} Fig.3:  Flow of time $\eta$ and a spatial variable $t$ in regions II and IV. (In this paper a flow of time $t$ in region III is flipped wrt the usual direction.)
       \end{center}
      \end{minipage}
\end{figure}
In region II $\eta$ is a time variable and the upper sign of (\ref{etapm}) will be used. It takes values in $\eta \geq \frac{2}{\sqrt{a}}$. In region IV the lower sign of (\ref{etapm}) is used and $\eta$ takes values in $0 \leq \eta \leq \frac{2}{\sqrt{a}}$. The variable $t$ is a spatial one in regions II and IV. In this paper region IV is not considered. 
One set of the normal modes  in region II obtained by a coordinate transformation from that for AdS$_3$ is given by (\ref{phiII2}),
\begin{equation}
\Phi^{\text{II}(\nu)}_{\omega, k}(\eta, t,x)
=   \frac{4+a\eta^2}{4\sqrt{a}\eta}\, e^{\sqrt{a}x}J_{\nu}(\mu \frac{4+a\eta^2}{4\sqrt{a}\eta}e^{\sqrt{a}x})\exp\{ i\frac{a\eta^2-4}{4\sqrt{a}\eta} e^{\sqrt{a}x}\mu \cosh (\zeta -\sqrt{a}t)\}, \label{phiII}
\end{equation}
where $\omega=\mu\cosh \zeta$ and $k=\mu\sinh \zeta$. 
Then the eigenfunctions of momenta corresponding to $t$ and $x$  translations are given by Fourier transformation of (\ref{phiII}) with respect to $\zeta$ and $\ln \mu$. 
\begin{multline}
\Phi^{\text{II}(\nu)}_{E,p}(\eta,t,x) = \int_{-\infty}^{\infty}d\zeta\int_0^{\infty}d\mu \, e^{-i\frac{E}{\sqrt{a}}\zeta-i\frac{p}{\sqrt{a}}\ln \mu} \, \Phi^{\text{II}}_{\omega,k}(\eta,t,x)  \\
\equiv e^{-iEt+ipx}\int_{-\infty}^{\infty}d\zeta\int_0^{\infty}d\mu e^{-i\frac{E}{\sqrt{a}}\zeta-i\frac{p}{\sqrt{a}}\ln \mu} 
\frac{4+a\eta^2}{4\sqrt{a}\eta}\, J_{\nu}(\mu \frac{4+a\eta^2}{4\sqrt{a}\eta})\exp\Big\{ i\frac{a\eta^2-4}{4\sqrt{a}\eta} \mu \cosh \zeta \Big\}.   \label{f}
\end{multline}
Here $E$ and $p$ are momentum eigenvalues conjugate to $t$ and $x$, and take values in $-\infty < E,p<\infty$. 
This solution can also be represented as a suitable linear combination of solutions constructed in terms of a hypergeometric functions as in \cite{Ichinose}\cite{Vakkuri}. 
This integration formula, however, allows explicit asymptotic formulas. 

This set of normal modes, however,  does not form a complete set of linearly independent functions. 
In (\ref{complexJ})  of Appendix C it is shown that the following relation holds for $\eta >2/\sqrt{a}$.
\begin{equation}
\big(\Phi^{\text{II}(\nu)}_{E,p} (\eta,t,x)\big)^{\ast} =e^{-\pi i (\nu+1)} \, e^{\beta p/2} \, \Phi^{\text{II}(\nu)}_{-E,-p}(\eta,t,x)  \label{complexJ}
\end{equation}
Let us note that $E$ is a component of spatial momentum. 
Then we will introduce  a new set of normal modes. 
\begin{multline}
\tilde{\Phi}^{\text{II}(\nu)}_{E,p}(\eta,t,x) 
= e^{-iEt+ipx}\int_{-\infty}^{\infty}d\zeta\int_0^{\infty}d\mu e^{-i\frac{E}{\sqrt{a}}\zeta+i\frac{p}{\sqrt{a}}\ln \mu} 
\frac{4+a\eta^2}{4\sqrt{a}\eta}\\ \times
 J_{\nu}(\mu \frac{4+a\eta^2}{4\sqrt{a}\eta})\exp\Big\{ i\frac{a\eta^2-4}{4\sqrt{a}\eta} \mu \cosh \zeta \Big\}.   \label{ftilde}
\end{multline}
This is related to (\ref{f}) by a relation\footnote{In the $(\omega,k)$ basis of (\ref{phiII}), a relation $\tilde{\Phi}^{\text{II}(\nu)}_{\omega,k}(\eta,t,x)=(\omega'^2-k'^2)\Phi^{\text{II}(\nu)}_{\omega',k'}(\eta,t,-x)$ with $(\omega',k')=(\omega/(\omega^2-k^2), k/(\omega^2-k^2))$ holds. }
\begin{equation}
\tilde{\Phi}^{\text{II}(\nu)}_{E,p}(\eta,t,x) =\Phi^{\text{II}(\nu)}_{E,-p}(\eta,t,-x). 
 \label{relII}
\end{equation}

The Klein-Gordon (K-G) inner products for these modes in region II, $\eta \geq 2/\sqrt{a}$,  is defined for functions $f(\eta,t,x)$ and $g(\eta,t,x)$ as follows.
\begin{equation}
(f,g)_{\text{II}} \equiv i\int_{-\infty}^{\infty}dt \int_{-\infty}^{\infty}dx\, \frac{a\eta (a\eta^2-4)}{a\eta^2+4}\Big[ f^{\ast}\partial_{\eta} g-g\partial_{\eta} f^{\ast}\Big] \label{KGinII}
\end{equation}
The factor in the integrand which depends on $\eta$ is $\sqrt{-g}(-g^{\eta\eta})$. The subscript II for the inner product on the left hand side means that the integration is to be carried out in region II. 
This inner product does not depend on $\eta$ owing to K-G equation (at least away from the horizon. See next section.) Hence the inner products of  the normal modes will be computed in the limit $\eta \rightarrow +\infty$. They are evaluated in Appendix C. Some of the results for $\eta >2/\sqrt{a}$ are
\begin{eqnarray}
(\Phi^{\text{II}(\nu)}_{E,p}, \Phi^{\text{II}(\nu)}_{E',p'})_{\text{II}}
&=&  \frac{-8\pi^3 a \exp (\beta p/2)}{ \sinh (\beta p/2)} \, 
\delta(E-E')\delta(p-p'), \label{innerII1} \\
(\tilde{\Phi}^{\text{II}(\nu)}_{E,p}, \tilde{\Phi}^{\text{II}(\nu)}_{E',p'})_{\text{II}}
&=&  \frac{8\pi^3 a \exp (-\beta p/2)}{ \sinh (\beta p/2)} \, 
\delta(E-E')\delta(p-p'), \label{innerII1tilde} \\
(\Phi^{\text{II}(\nu)}_{E,p}, \tilde{\Phi}^{\text{II}(\nu)}_{E',p'})_{\text{II}}
&=&  0
\end{eqnarray}

These inner products are not positive definite: for example, (\ref{innerII1}) is negative for $p>0$ and positive for $p<0$. Furthermore,  (\ref{f}) behaves as $\Phi^{\text{II}(\nu)}_{E,p} \propto \eta^{ip/\sqrt{a}} \exp (-iEt+ipx)$ at large $\eta$.  Similarly (\ref{ftilde}) behaves as $\tilde{\Phi}^{\text{II}(\nu)}_{E,p} \propto \eta^{-ip/\sqrt{a}} \exp (-iEt+ipx)$. By using these properties of mode functions  it can be shown that 
\begin{equation}
(\Phi^{\text{II}(\nu)\ast}_{E,-p}, \tilde{\Phi}^{\text{II}(\nu)}_{-E',p'})_{\text{II}}=0 \label{innerII2}
\end{equation}
for $-\infty <E, E',p,p'<\infty$. 

In addition to $\Phi^{\text{II}(\nu)}_{E,p}$ and $\tilde{\Phi}^{\text{II}(\nu)}_{E,p}$ we will also  introduce new normal modes, or basis functions, $\Phi^{\text{II}(-\nu)}_{E,p}$ and $\tilde{\Phi}^{\text{II}(-\nu)}_{E,p}$, which are obtained from 
 $\Phi^{\text{II}(\nu)}_{E,p}$ and $\tilde{\Phi}^{\text{II}(\nu)}_{E,p}$, respectively, by replacing Bessel function $J_{\nu}(z)$ by $J_{-\nu}(z)$. 
So there are four types of basis functions: 
\begin{equation}
{\cal V} =\big\{\Phi^{\text{II}(\nu)}_{E,p},  \quad \Phi^{\text{II}(-\nu)}_{E,p}, \quad \tilde{\Phi}^{\text{II}(\nu)}_{E,p}, \quad \tilde{\Phi}^{\text{II}(-\nu)}_{E,p} \big\}  \label{fourmodes}
\end{equation}
Although usually, only two of these functions are chosen to  be a basis of linearly independent functions, we will consider a vector space ${\cal V}$ of functions spanned by this set of the normal modes, and  expand the scalar field inside horizon  into these modes. This is because in curved spacetime without time-translation symmetry positive-and negative-frequency solutions cannot be defined and it is not known beforehand which normal modes are to be assigned to annihilation operators. So the number of independent basis functions must be doubled. Another reason is that it will be shown in the next section that the inner products of these functions are not independent of $\eta$  at the horizon, {\em i.e.,} take distinct values at the horizon from those for $\eta> 2/\sqrt{a}$. Therefore it is necessary to keep all of the modes (\ref{fourmodes}) as the basis in the analysis of matching conditions at the horizon. Furthermore, $\Phi^{\text{II}(\nu)}_{E,p}$ and 
$\Phi^{\text{II}(\nu)\ast }_{-E,-p}$ are not linearly independent as shown above.
So the mode functions with $p >0$ and those with $p<0$ must be treated separately. 
The above four mode functions (\ref{fourmodes}) are all proportional to $e^{-iEt+ipx}$. 

The scalar field must satisfy matching conditions at the horizon\cite{Raju1} as well as  normalization condition of the inner products of the normal modes. For this prescription to work the inner products of the normal mode functions which are used to expand the scalar field must be constructed in such a way that they are smooth in the interior region of the black hole including the horizon.  

\section{Discontinuity of the Inner Product in $\eta$ at the Horizon}
\hspace*{5mm}
In this section it will be shown that although the inner product (\ref{KGinII}) of the normal modes is independent of time $\eta$ for $\eta > 2/\sqrt{a}$, it will change discontinuously at $\eta=2/\sqrt{a}$. Only inner products of appropriate linear combinations of the normal modes will be continuous at the horizon. 

If $f$ and $g$ are solutions to K-G equation, the inner product (\ref{KGinII}) satisfies the following equation owing to the K-G equation.
\begin{eqnarray}
\frac{d}{d\eta}(f,g)_{\text{II}}&=&i\int dt\int dx \partial_t \big[ g^{tt}\sqrt{-g}(f^{\ast}\partial_tg-g\partial_t f^{\ast})\big] \nonumber \\
&& +i\int dt\int dx \partial_x \big[ g^{xx}\sqrt{-g}(f^{\ast}\partial_xg-g\partial_x f^{\ast})\big]
\label{conservation}
\end{eqnarray}
Here the factor $g^{tt}\sqrt{-g}$ in the first line is given by
\begin{equation}
g^{tt}\sqrt{-g}= \frac{16a\eta}{(a\eta^2+4)(a\eta^2-4)}
\end{equation}
This has a singularity exactly at the location of the horizon $\eta=2/\sqrt{a}$. Because the integrand is a total derivative, the right hand side will vanish for $\eta >2/\sqrt{a}$. Then the inner products are independent of $\eta$ as long as $\eta >2/\sqrt{a}$.  
Just at the horizon, however, special care must be taken. This problem can be studied by computing the inner product of the mode functions as $\eta \rightarrow \infty$ and at $\eta=2/\sqrt{a}$, separately, and showing that the results of the two cases are same or distinct.\footnote{In the case of Rindler space it can be shown that there is also a singularity at the horizon in an equation similar to (\ref{conservation}). In this case, however, inner products of mode functions are not discontinuous at the horizon. }

Inner products of (\ref{fourmodes}) at $\eta >2/\sqrt{a}$ are given in  (\ref{table1})-(\ref{table8}).  These are computed by using the asymptotic behaviors (\ref{limitPII}) for $\eta \rightarrow \infty$.   Inner products of the four modes in (\ref{fourmodes}) can also be computed in a region near the horizon. In (\ref{nearhorII})  a behavior of $\Phi^{\text{II}(\nu)}_{E,p}$ near the horizon is presented. By using this result some inner products of (\ref{fourmodes}) are computed in Appendix D. The results are completely different from those in (\ref{table1})-(\ref{table8}).  Hence the inner products of the mode functions have discontinuities at $\eta=2/\sqrt{a}$. 

Next, we will study linear dependence of the mode functions (\ref{fourmodes}). In the $\eta \rightarrow \infty$ limit, $\Phi^{\text{II}(\nu)}_{E,p}$ behaves as (\ref{limitPII}). 
This shows that there exist following two relations in this limit. 
\begin{eqnarray}
&& \Phi^{\text{II}(-\nu)}_{E,p} - e^{-\pi i \nu}D(E,p)\Phi^{\text{II}(\nu)}_{E,p}=0, \label{dep1}\\
&& \tilde{\Phi}^{\text{II}(-\nu)}_{E,p} - e^{-\pi i \nu}D^{\ast}(E,p)\tilde{\Phi}^{\text{II}(\nu)}_{E,p}=0  \label{dep2}
\end{eqnarray}  
Here $D(E,p)$ is defined in (\ref{DEp}).  It can be shown that inner products of the left hand sides of (\ref{dep1}) and (\ref{dep2}) with the four functions in (\ref{fourmodes}) all vanish, as long as $\eta >2/\sqrt{a}$. These relations are valid for $\eta >2/\sqrt{a}$.   There are only two linearly independent modes, as should be the case. 

On the contrary by using the behavior (\ref{nearhorII}) of $\Phi^{\text{II}(\nu)}_{E,p}$ near the horizon,  it can be shown that the following relation holds near the  horizon.
\begin{equation}
 \tilde{\Phi}^{\text{II}(\nu)}_{E,p}=e^{2ip/\sqrt{a}}\frac{\Gamma(\frac{1}{2}(1+\nu+i\frac{E+p}{\sqrt{a}}))\Gamma(\frac{1}{2}(1+\nu-i\frac{E-p}{\sqrt{a}}))       }{ \Gamma(\frac{1}{2}(1+\nu-i\frac{E+p}{\sqrt{a}})) \Gamma(\frac{1}{2}(1+\nu+i\frac{E-p}{\sqrt{a}}))  } \Phi^{\text{II}(\nu)}_{E,p}  \label{relhorizon}
\end{equation} 
Another  relation obtained by a replacement $\nu \rightarrow -\nu$ also holds. 
In this way linear dependence of mode functions may be discontinuous at the horizon. 
It is then unavoidable to use all the four modes altogether behind the horizon.
So we will work in the vector space of normal modes ${\cal V}$. 

Now a question arises as to how it is possible to match a scalar field in regions I and III outside the horizon to that in region II. Even if it could be possible to connect the mode functions of a scalar field on both sides of the horizon smoothly, the inner products of the mode functions might be discontinuous at the horizon $\eta=2/\sqrt{a}$. How the creation and annihilation operators must be assigned to the mode functions inside region II? 
If the inner products at $\eta=2/\sqrt{a}$ were used, the quantum theory obtained might not be appropriate in the whole region inside the horizon. If the inner products for $\eta > 2/\sqrt{a}$ were used, then the quantum theory inside the horizon might be discontinuous from that outside. In the following analysis we will start by  using the inner products for $\eta >2/\sqrt{a}$, (\ref{table1})-(\ref{table8}) and solve the matching conditions. It will be shown in sec.8 that when these conditions are satisfied, then special linear combinations of normal modes, $\psi^{(i)}_{E,p}$ (\ref{psi1234}) defined below,  have appropriate inner products which are continuous at the horizon. 

\section{Quantization Behind the Horizon}
\hspace*{5mm}
Behind the horizon the scalar field will be expanded into linear combinations of the normal modes  (\ref{fourmodes})  introduced above. For the horizon to be smooth, the operators which multiply the normal modes must be chosen to coincide with those in regions I and III, and the operators inside and outside the horizon must match appropriately. In this section the matching condition for normal modes will be studied and solutions to the condition will be obtained. It is found that the solution contains parameters $F^{(1)}_n(E,p)$, $F^{(2)}_n(E,p)$ ($n=1,2)$ and it is not unique. 

 \subsection{Normal Mode Expansion  of a Scalar Field in Region II}
\hspace*{5mm}
Choice of normal modes is carried out in such a way that the exponentials $e^{\pm iEt \pm ipx}$ in the normal modes in region II coincide with those in the corresponding normal modes in regions I and III, respectively. 
Normal mode expansion of a scalar field in region II is given by
\begin{multline}
\Phi^{\text{II}}(\eta,t,x) = \frac{1}{4\pi 
\sqrt{\pi a}}\int_0^{\infty}dE \int_0^{\infty}dp \\ 
  \Big\{c_R(E,-p) \psi^{(1)}_{E,-p}+ c_R(E,p) \psi^{(2)}_{E,p} \\ 
+ c_L(E,-p)\psi^{(3)}_{E,-p} +c_L(E,p) \psi^{(4)}_{E,p}  +h.c. \Big\},
\label{PhiII2}
\end{multline}
Here $c_R(E,p)$ and $c_L(E,p)$ with $E \geq 0$ and $-\infty <p<+\infty$ are two sets of annihilation operators which are the same as those in regions I and III. It is important for smoothness of the horizon to use the same operators inside the horizon as those outside.\cite{Raju1} 
In the above expansion of a scalar field,   integration region for $p$ is devided into two, $p > 0$ and $p<0$, because $\Phi^{\text{II}(\nu)}_{E,p}$ and $\Phi^{\text{II}(\nu)}_{-E,-p}$ 
are related to each other by complex conjugation.  
$\psi^{(i)}_{E,p}(\eta,t,x) $ are normal mode functions defined by
\begin{multline}
\psi^{(1)}_{E,-p} = N^{(1)}_{E,p}[\gamma^{(1)}_1(E,p) \Phi^{\text{II}(\nu)}_{E,-p}+\gamma^{(1)}_2(E,p)\tilde{\Phi}^{\text{II}(\nu)}_{E,-p}  + \gamma^{(1)}_3(E,p) \Phi^{\text{II}(-\nu)}_{E,-p}+\gamma^{(1)}_4(E,p)\tilde{\Phi}^{\text{II}(-\nu)}_{E,-p}       ], \\
\psi^{(2)}_{E,p} = N^{(2)}_{E,p} [\gamma^{(2)}_1(E,p) \tilde{\Phi}^{\text{II}(\nu)}_{E,p}+\gamma^{(2)}_2(E,p)\Phi^{\text{II}(\nu)}_{E,p}  + \gamma^{(2)}_3(E,p) \tilde{\Phi}^{\text{II}(-\nu)}_{E,p}+\gamma^{(2)}_4(E,p)\Phi^{\text{II}(-\nu)}_{E,p}       ],    \\
 \psi^{(3)}_{E,-p} = N^{(3)}_{E,p}[\gamma^{(3)}_1(E,p) \tilde{\Phi}^{\text{II}(\nu)}_{-E,p}+\gamma^{(3)}_2(E,p)\Phi^{\text{II}(\nu)}_{-E,p}  + \gamma^{(3)}_3(E,p) \tilde{\Phi}^{\text{II}(-\nu)}_{-E,p}+\gamma^{(3)}_4(E,p)\Phi^{\text{II}(-\nu)}_{-E,p}       ], \\
 \psi^{(4)}_{E,p} = N^{(4)}_{E,p}[\gamma^{(4)}_1(E,p) \Phi^{\text{II}(\nu)}_{-E,-p}+\gamma^{(4)}_2(E,p)\tilde{\Phi}^{\text{II}(\nu)}_{-E,-p}  + \gamma^{(4)}_3(E,p) \Phi^{\text{II}(-\nu)}_{-E,-p}+\gamma^{(4)}_4(E,p)\tilde{\Phi}^{\text{II}(-\nu)}_{-E,-p}       ],  \label{psi1234}
\end{multline}
where  $E \geq 0$ and $p \geq 0$. If only $\Phi^{\text{II}(\nu)}_{E,p}$ and $\Phi^{\text{II}(-\nu)}_{E,p}$ are used in the definition of the normal modes,  it is not possible to impose matching conditions of the mode function $\psi^{(i)}_{E,p}$ at the horizon in addition to the normalization of the inner products. 
Functions $N^{(i)}_{E,p}$ and $\gamma^{(i)}_n(E, p)$ ($i=1,2$) should be chosen to make $\psi^{(i)}_{E,\mp p}$ behave as $(\sqrt{a}\eta -2)^{-iE/\sqrt{a}}$ and match $\psi^{(1)}_{E,- p}$ and $\psi^{(2)}_{E, p}$   smoothly with $\sqrt{1-e^{-\beta E}}\Phi^{\text{I}}_{E,-p}$ and $\sqrt{1-e^{-\beta E}}\Phi^{\text{I}}_{E,p}$, respectively, on the horizon between regions I and II. 
Similarly, $N^{(i)}_{E,p}$ and $\gamma^{(i)}_{n}(E, p)$ ($i=3,4$) are chosen to make $\psi^{(i)}_{E, \pm p}$  behave as $(\sqrt{a}\eta -2)^{+iE/\sqrt{a}}$ and match $\psi^{(4)}_{E, p}$ and $\psi^{(3)}_{E,- p}$ smoothly with $\sqrt{1-e^{-\beta E}}\Phi^{\text{III}}_{-E,-p}$ and $\sqrt{1-e^{-\beta E}}\Phi^{\text{III}}_{-E,p}$, respectively, on the horizon between regions II and III. 

It can be shown that in order to ensure that  $(\psi^{(3)\ast}_{E, -p},\psi^{(1)}_{E', -p'})=0$  it is necessary and sufficient to require $\gamma^{(3)}_1/\gamma^{(1)}_1=\gamma^{(3)}_2/\gamma^{(1)}_2$ and $\gamma^{(3)}_3/\gamma^{(1)}_3=\gamma^{(3)}_4/\gamma^{(1)}_4$. 
In the following the following constraints are imposed.
\begin{equation}
\gamma^{(3)}_n(E,p)= \gamma^{(1)}_n(E,p) \qquad (n=1,2,3,4)
\end{equation}
Although these constraints are a bit stronger than necessary, it is possible to show that there exist solutions to the matching conditions. Similarly, in order to ensure that  $(\psi^{(4)\ast}_{E, p},\psi^{(2)}_{E', p'})=0$ the following constraints are imposed.
\begin{equation}
\gamma^{(4)}_n(E,p)= \gamma^{(2)}_n(E,p) \qquad (n=1,2,3,4)
\end{equation}
On the other hand orthogonality of $\psi^{(i)}_{E,\pm p}$ and $\psi^{(j)}_{ E', \pm p'}$ with $i \neq j$ is ensured by the orthogonality properties of $e^{\pm iEt \pm i px}$.

The operators $c_{R,L}(E, \mp p)$ are assigned to each normal mode functions in such a way that they will correspond to appropriate $t$- and $x$-dependences of the normal modes multiplying $c_R$ and $c_L$ in $\Phi^{\text{I}}$ and $\Phi^{\text{III}}$, {\em i.e.}, $e^{\pm iEt \pm ipx}$, in (\ref{PhiR4}) and (\ref{PhiL4}).  The assignment of these normal modes in (\ref{PhiII2}) are carried out by taking into account (\ref{innerII2}). 
If the normal modes $\psi^{(i)}_{E,\mp p }$ defined as above have  `norm's  normalized to unity as
\begin{equation}
(\psi^{(i)}_{ E, \pm p}, \psi^{(i)}_{ E', \pm p'})_{\text{II}} = 16\pi ^3 a\delta(E-E')\delta(p-p'), \qquad (i=1,2,3,4)  \label{wouldbe}
\end{equation}
then  it is possible to multiply the normal modes $\psi^{(i)}_{E,\mp p }$ by operators $c_R(E,\pm p)$ and $c_L(E,\pm p)$ as in (\ref{PhiII2}). 

The norms of the normal modes $\psi^{(i)}_{E,\mp p}$ $(i=1, \cdots, 4)$ are computed by using the results in Appendix B as
\begin{multline}
(\psi^{(i)}_{E, \mp p},\psi^{(i)}_{E', \mp p'})_{\text{II}}
=8\pi^3 a[\sinh (\beta p/2)]^{-1} |N_{E,p}^{(i)}|^2\delta(E-E')\delta(p-p') \\ \times 
\Big[e^{-\beta p/2}|\gamma^{(i)}_1+e^{-\pi i\nu}D^{\ast}(E,p)\gamma^{(i)}_3|^2-e^{\beta p/2}|\gamma^{(i)}_2+e^{-\pi i\nu}D(E,p)\gamma^{(i)}_4|^2 \Big]  \label{norm1}
\end{multline}
Although in general these norms are not positive definite, by imposing $\gamma^{(i)}_2=\gamma^{(i)}_4=0$, (\ref{norm1}) could be made positive definite. In this case, however, a solution to the  matching conditions of the modes at the horizon would not satisfy appropriate normalization conditions of the normal modes. Another prescription for restriction, $\gamma^{(i)}_3=\gamma^{(i)}_4=0$, does not work, either.  In what follows, suitable solutions which have positive norms will be found. 

\subsection{Matching Conditions}
\hspace*{5mm}
In the maximally extended Penrose diagram the black hole spacetime is expressed in terms of the Kruskal-Szeckeres coordinates, $U$ and $V$.  Each regions have their own $U$, $V$ coordinates.
\begin{figure}[thb]
\hspace {4cm}     \begin{minipage}{0.5\hsize}
      \begin{center}
\includegraphics[scale=.3, clip]{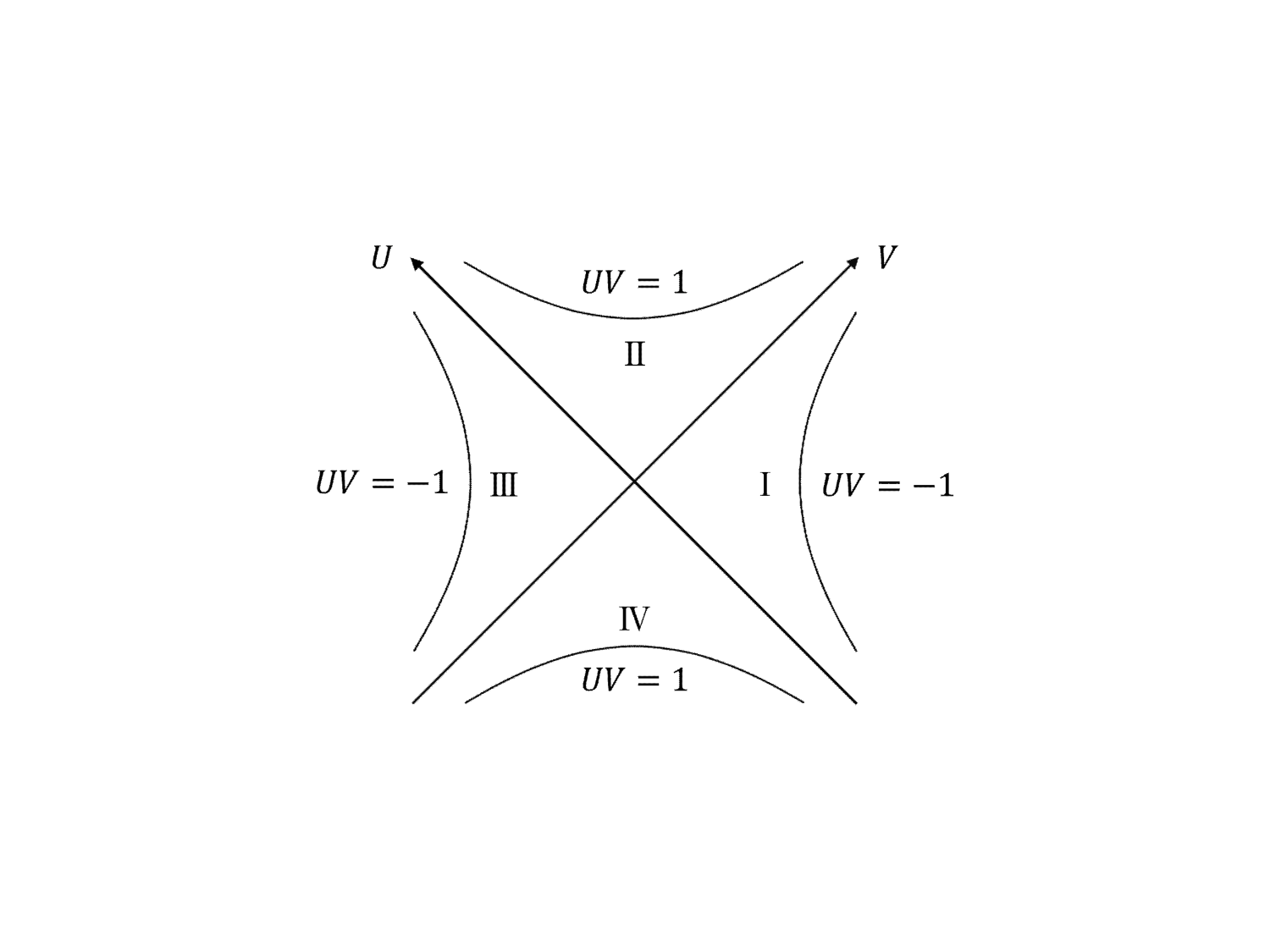}
\hspace{1.8cm} Fig. 4: Kruskal-Szeckeres coordinates. 
        \end{center}
      \end{minipage}
\end{figure}
In region I these coordinates are given by 
\begin{equation}
U_{\text{I}} =-e^{\sqrt{a}(r_{\ast}-t)  }= -e^{-\sqrt{a}t} \frac{2-\sqrt{a}y}{ 2+\sqrt{a}y}, \qquad V_{\text{I}}=e^{\sqrt{a}(r_{\ast}+t)  }=e^{\sqrt{a}t} \frac{2-\sqrt{a}y}{ 2+\sqrt{a}y}. 
\end{equation} 
Here $r_{\ast}$ is the tortoise coordinate. 
In region II they are given by 
\begin{equation}
U_{\text{II}} = e^{-\sqrt{a}t} \frac{\sqrt{a}\eta-2}{ \sqrt{a}\eta+2}, \qquad V_{\text{II}}=e^{\sqrt{a}t} \frac{\sqrt{a}\eta-2}{ \sqrt{a}\eta+2}.
\end{equation} 
Similarly, in region III they are 
\begin{equation}
U_{\text{III}} = e^{\sqrt{a}t} \frac{\sqrt{a}y-2}{ \sqrt{a}y+2}, \qquad V_{\text{III}}=-e^{-\sqrt{a}t} \frac{\sqrt{a}y-2}{ \sqrt{a}y+2}. 
\end{equation} 
Here the time direction in region III is flipped compared to that in the usual maximally-extended Penrose diagram to make the positive direction of $t$ upward. 
Then the near-horizon behaviors of normal modes are either $V^{\pm iE/\sqrt{a}}$ or $U^{\pm iE/\sqrt{a}}$. Near the horizon between regions I and II, the coefficients $\gamma^{(1)}_n$ and $\gamma^{(2)}_n$  in (\ref{psi1234}) must be determined to keep only terms proportional to $V^{- iE/\sqrt{a}}$ in region II. Furthermore, the terms proportional to $V^{- iE/\sqrt{a}}$ on both sides of the horizon must coincide. This determines $N^{(1,2)}_{E,p} \gamma^{(1,2)}_n(E,p)$ ($n=1, \cdots ,4)$. Similarly near the horizon between regions III and II,  $\gamma^{(3)}_n(=\gamma^{(1)}_n)$ and $\gamma^{(4)}_n(=\gamma^{(2)}_n)$ must be determined to make the terms on both sides of the horizon proportional to $U^ {iE/\sqrt{a}}$, and it can be shown that the requirement that the terms proportional to $(-V)^{-iE/\sqrt{a}}$ in region II near the horizon should vanish is also satisfied. In the following this analysis is sketched. 

The near-horizon behavior of $\Phi^{\text{I}}_{E,p}$, $\Phi^{\text{III}}_{E,p}$, $\Phi^{\text{II}(\pm \nu)}_{E,p}$ and $\tilde{\Phi}^{\text{II}(\pm \nu)}_{E,p}$ are presented in Appendix C. By using these formulas the matching conditions  at the horizon for the scalar field can be examined.   Then, it can be checked whether the inner products of the normal modes satisfy (\ref{wouldbe}). 

The matching conditions are given as follows.
\begin{itemize}
\item At the horizon between regions I and II  ($U=0, V>0$)
\begin{eqnarray}
\psi^{(1)}_{E,-p}&=&N^{(1)}_{E,p}e^{-ipx}[\gamma^{(1)}_1C^{\nu}(E,-p)+\gamma^{(1)}_2C^{\nu}(E,p) \nonumber \\
&& +\gamma^{(1)}_3C^{-\nu}(E,-p)+\gamma^{(1)}_4C^{-\nu}(E,p)]V^{-iE/\sqrt{a}} \nonumber \\
&=& \sqrt{1-e^{-\beta E}}e^{-ipx}2^{ip/\sqrt{a}}\frac{\Gamma(\frac{1+\nu}{2}-i\frac{E-p}{2\sqrt{a}})}{ \Gamma(\frac{1+\nu}{2}+i\frac{E-p}{2\sqrt{a}}) } e^{\beta E/4}
\Gamma(i\frac{E}{\sqrt{a}})(2V)^{-iE/\sqrt{a}} \nonumber \\
&=& b_1 \, e^{-ipx} V^{-iE/\sqrt{a}},  \label{matching1} \\
\psi^{(2)}_{E,p}&=&N^{(2)}_{E,p}e^{ipx}[\gamma^{(2)}_1C^{\nu}(E,-p)+\gamma^{(2)}_2C^{\nu}(E,p)\nonumber \\
&&+\gamma^{(2)}_3C^{-\nu}(E,-p)+\gamma^{(2)}_4 C^{-\nu}(E,p)]V^{-iE/\sqrt{a}} \nonumber \\
&=& \sqrt{1-e^{-\beta E}}e^{ipx}2^{-ip/\sqrt{a}}\frac{\Gamma(\frac{1+\nu}{2}-i\frac{E+p}{2\sqrt{a}})}{ \Gamma(\frac{1+\nu}{2}+i\frac{E+p}{2\sqrt{a}}) } e^{\beta E/4}
\Gamma(i\frac{E}{\sqrt{a}})(2V)^{-iE/\sqrt{a}} \nonumber \\
&=& b_2 \, e^{ipx} V^{-iE/\sqrt{a}}.  \label{matching2}
\end{eqnarray}
\item At the horizon  between regions II and III ($V=0, U>0$ )
\begin{eqnarray}
\psi^{(3)}_{E,-p}&=&N^{(3)}_{E,p}e^{ipx}[\gamma^{(1)}_1C^{\nu}(E,-p)+\gamma^{(1)}_2 C^{\nu}(E,p)\nonumber \\
&& +\gamma^{(1)}_3C^{-\nu}(E,-p)+\gamma^{(1)}_4C^{-\nu}(E,p)]U^{-iE/\sqrt{a}} \nonumber \\
&=& \sqrt{1-e^{-\beta E}}e^{ipx}2^{-ip/\sqrt{a}}\frac{\Gamma(\frac{1+\nu}{2}-i\frac{E+p}{2\sqrt{a}})}{ \Gamma(\frac{1+\nu}{2}+i\frac{E+p}{2\sqrt{a}}) } e^{\beta E/4}
\Gamma(i\frac{E}{\sqrt{a}})(2U)^{-iE/\sqrt{a}} \nonumber \\
&=& b_3 \, e^{ipx} U^{-iE/\sqrt{a}}, \label{matching3} \\
\psi^{(4)}_{E,p}&=&N^{(4)}_{E,p}e^{-ipx}[\gamma^{(2)}_1C^{\nu}(E,-p)+\gamma^{(2)}_2C^{\nu}(E,p) \nonumber \\
&& +\gamma^{(2)}_3C^{-\nu}(E,-p)+\gamma^{(2)}_4C^{-\nu}(E,p)]U^{-iE/\sqrt{a}} \nonumber \\
&=& \sqrt{1-e^{-\beta E}}e^{-ipx}2^{ip/\sqrt{a}}\frac{\Gamma(\frac{1+\nu}{2}-i\frac{E-p}{2\sqrt{a}})}{ \Gamma(\frac{1+\nu}{2}+i\frac{E-p}{2\sqrt{a}}) } e^{\beta E/4}
\Gamma(i\frac{E}{\sqrt{a}})(2U)^{-iE/\sqrt{a}} \nonumber \\
&=& b_4 \, e^{-ipx} U^{-iE/\sqrt{a}}  \label{matching4}
\end{eqnarray}
\end{itemize}
The conditions for vanishing of terms in $\psi^{(1,2)}$ proportional to $(-U)^{iE/\sqrt{a}}$ at the horizon are given by
\begin{eqnarray}
&\gamma^{(1)}_1 C^{\nu}(-E,-p)+\gamma^{(1)}_2C^{\nu}(-E,p)+\gamma^{(1)}_3 C^{-\nu}(-E,-p)+\gamma^{(1)}_4C^{-\nu}(-E,p)=0, \label{matching5} \\
&\gamma^{(2)}_1 C^{\nu}(-E,-p)+\gamma^{(2)}_2 C^{\nu}(-E,p)+\gamma^{(2)}_3 C^{-\nu}(-E,-p)+\gamma^{(2)}_4 C^{-\nu}(-E,p)=0 \label{matching6}
\end{eqnarray}
Exactly the same conditions are obtained for vanishing of terms in $\psi^{(3,4)}$ proportional to $(-V)^{iE/\sqrt{a}}$ at the horizon.
Here $C^{\pm \nu}$ are defined by 
\begin{equation}
C^{\pm \nu}(E,p)= e^{-\beta E/4}\Gamma(i\frac{E}{\sqrt{a}}) 2^{-i\frac{E+p}{\sqrt{a}}} 
\frac{\Gamma(\frac{1 \pm \nu}{2}-i\frac{E+p}{2\sqrt{a}})}{ \Gamma(\frac{1\pm \nu}{2}+i\frac{E+p}{2\sqrt{a}}) } 
\end{equation}
Then $\psi^{(3)}_{E,-p}$ and $ \psi^{(4)}_{E,p}$ are proportional to $U^{-iE/\sqrt{a}}$ and vanish at the horizon between, where $U=0$. Similarly, $\psi^{(1)}_{E,-p}$ and $\psi^{(2)}_{E,p}$ are proportional to $V^{-iE/\sqrt{a}}$ and vanish at the horizon between III and II.

\subsection{Solutions}
\hspace*{5mm}
By solving (\ref{matching1})-(\ref{matching6}) the following solution for $\gamma^{(i)}_n$'s are obtained. Details are given in Appendix F. 
\begin{eqnarray}
\gamma^{(1)}_1 &= & K\, e^{-\pi i\nu} [e^{\beta p/2} e^{i\delta_1}-D^{\ast}(E,p)(e^{-\beta p/2}e^{i\delta_2}-J^{(1)})], \label{gamma1}\\
\gamma^{(1)}_3 &=& K \, [-e^{\beta p/2} e^{i\delta_1}+e^{-\beta p/2}e^{i\delta_2}-J^{(1)}], \label{gamma3}
\end{eqnarray}
Here 
\begin{eqnarray}
K & \equiv & \frac{[e^{\beta (E-p)/2}+e^{\pi i\nu}][e^{\beta(E+p)/2}+e^{-\pi i \nu}]}
{4i e^{\beta E/2} \, \sinh(\beta p/2) e^{-\pi i \nu} \, \sin \pi \nu}, \\
J^{(1)} & \equiv & \frac{b_1}{N^{(1)}_{E,p}}\frac{Ee^{-\beta E/2}e^{-\pi i \nu}}{2\pi \sqrt{a}} C^{\nu}(-E,p)[e^{\beta (E-p)/2}+e^{\pi i \nu}] \label{J1}
\end{eqnarray}
$b_1$ is defined in (\ref{b1}), and 
\begin{eqnarray}
\gamma^{(1)}_2 &=& -\frac{M_1}{M_2} \, \gamma^{(1)}_1+\frac{C^{-\nu}(-E,p)}{M_2(E,p)}\frac{b_1}{N^{(1)}_{E,p}}, \label{gamma2}\\
\gamma^{(1)}_4 &=& \frac{M_3}{M_2}\, \gamma^{(1)}_3-\frac{C^{\nu}(-E,p)}{M_2(E,p)}\frac{b_1}{N^{(1)}_{E,p}}, \label{gamma4}
\end{eqnarray}
Here $M_{1,2,3}$ are defined in (\ref{M1})-(\ref{M3}). 
Solutions for $\gamma^{(2)}_n$ with $n=1,2,3,4$ are also presented in Appendix E. 

By using these results norms of $\psi^{(i)}_{E, \mp p}$($i=1, \cdots,4)$ are found to be 
\begin{equation}
(\psi^{(i)}_{E, \mp p},\psi^{(i)}_{E',\mp p'})_{\text{II}}= 16\pi^3 a\, |N^{(i)}_{E,p}|^2 \delta(E-E') \delta (p-p'). \label{correctnorm}
\end{equation}
If $ N^{(i)}_{E,p} $'s are phase factors, (\ref{wouldbe}) hold. 
These norms are positive-definite and normalized to unity. Due to the pre-factor $1/(4\pi \sqrt{\pi a})$ in the scalar field (\ref{PhiII2}),  
the operators $c_{R,L}(E,p)$ must satisfy 
\begin{equation}
[c_R(E,p), c^{\dagger}_R(E',p')]=  \delta(E-E')\delta(p-p')
\end{equation}
and a similar relation for $c_L$ and $c_L^{\dagger}$. These relations agree with (\ref{FourierR}) and (\ref{FourierL}) and  the horizon is  smooth for the scalar field.  The prescription of \cite{Raju3} works and there are no firewalls for a scalar at the horizon. 

The above solution, however, contains phase factors $e^{i\delta_i(E,p)}$ $(i=1, \cdots,4)$, $N^{(1-4)}_{E,p}$. 
Furthermore, it is possible to obtain more general solutions which satisfy the matching conditions and the condition of normalization of normal modes. Let $F_1^{(i)}(E,p)$ and $F_2^{(i)}(E,p)$ ($i=1,2$) be real functions of $E$ and $p$ which satisfy for $E,p>0$ 
\begin{eqnarray}
&&F_1^{(i)}(E,p), \  F_2^{(i)}(E,p) \geq 0 \qquad \ (i=1,2), \\
&&e^{-\beta p/2}F_1^{(i)}(E,p) - e^{\beta p/2} F_2^{(i)}(E,p) =e^{\beta p/2}-e^{-\beta p/2} \qquad (i=1,2) \label{FF}
\end{eqnarray}
These functions $F^{(i)}_n$ can be parametrized as
\begin{eqnarray}
F^{(i)}_1(E,p) &=& (e^{\beta p}-1) \, \cosh^2 \alpha^{(i)}(E,p)+e^{\beta p/2}W^{(i)}(E,p), \label{W1} \\
F^{(i)}_2(E,p) &=& (1-e^{-\beta p}) \, \sinh^2 \alpha^{(i)}(E,p)+e^{- \beta p/2}W^{(i)}(E,p), \label{W2}
\end{eqnarray}
where $\alpha^{(i)}(E,p) \geq 0$ and $W^{(i)}(E,p) \geq 0$.  ($i=1,2$)

Then $\gamma^{(1)}_{1}$ and $\gamma^{(1)}_{3}$ are defined to be solutions to the following equations
\begin{eqnarray}
& |\gamma^{(1)}_1+\gamma^{(1)}_3 e^{-\pi i \nu} D^{\ast}(E,p)|^2 = F^{(1)}_1(E,p), \label{asd1}\\
& |\gamma^{(1)}_1+\gamma^{(1)}_3 e^{-\pi i \nu} +J^{(1)}|^2 = F^{(1)}_2(E,p) \label{asd2}
\end{eqnarray}
If $F^{(1)}_1=e^{\beta p}$, $F^{(1)}_2=e^{-\beta p}$, these agree with the results mentioned above. 
In the general  case the phase factors $e^{i\delta_{1,2}}$ in the above results are replaced by $e^{i\delta_{1,2}}\sqrt{F^{(1)}_{1,2}}$. 
In the same way by defining $\gamma^{(2)}_{1,3}$ in terms of $F^{(2)}_{1,2}(E,p)$ as in (\ref{asd1}) and (\ref{asd2}) the solutions for $\gamma^{(2)}_n$ ($n=1, \cdots, 4$) are obtained. 
These solutions depend on phase factors $e^{i\delta_{1,2,3,4}}$ and functions $F^{(1,2)}_{1,2}(E,p)$. So there are too many possibilities for quantum theory of the scalar field inside the horizon. 

\section{Inner Products of $\psi^{(i)}_{E,p}$ at the Horizon}
\hspace*{5mm}
In sec. 6 it was noticed that the inner products of the basis functions (\ref{fourmodes}) are discontinuous in $\eta$ at $\eta=2/\sqrt{a}$. In this section it will be studied whether the inner products of $\psi^{(i)}_{E,p}$ in (\ref{psi1234}) are continuous or not, if the solutions for $\gamma^{(i)}_n$ are substituted into (\ref{psi1234}). It will be shown that for the solution to the matching condition the inner products of the normal modes at the horizon is smoothly connected to those far from the horizon. 

Let us consider $\psi^{(1)}_{E,-p}$. By substitution of (\ref{gamma2}) and (\ref{gamma4}) it is found that
\begin{eqnarray}
\psi^{(1)}_{E,-p} &=& N^{(1)}_{E,p} \, \gamma^{(1)}_1 \big\{\Phi^{\text{II}(\nu)}_{E,-p}-\frac{M_1}{M_2}\tilde{\Phi}^{\text{II}(\nu)}_{E,-p}\big\}+N^{(1)}_{E,p} \, \gamma^{(1)}_3 \big\{\Phi^{\text{II}(-\nu)}_{E,-p}+\frac{M_3}{M_2}\tilde{\Phi}^{\text{II}(-\nu)}_{E,-p}\big\} \nonumber \\
&&+ \frac{b_1}{M_2 } \, C^{-\nu}(-E,p) \big\{ \tilde{\Phi}^{\text{II}(\nu)}_{E,-p}-\frac{C^{\nu}(-E,p)}{C^{-\nu}(-E,p)} \, \tilde{\Phi}^{\text{II}(-\nu)}_{E,-p}\big\} \label{psi1horizon}
\end{eqnarray}
At the horizon those factors in the first line which multiply $\gamma^{(1)}_1$ and $\gamma^{(1)}_3$, respectively, vanish due to the relation (\ref{relhorizon}).  Actually, the latter relation can be rewritten as $\Phi^{\text{II}(\nu)}_{E,-p}= [C^{\nu}(-E,-p)/C^{\nu}(-E,p)]\, \tilde{\Phi}^{\text{II}(\nu)}_{E,-p} =[M_1/M_2] \, \tilde{\Phi}^{\text{II}(\nu)}_{E,-p}$. So $\psi^{(1)}_{E,-p}$ is independent of $\gamma^{(1)}_{E,-p}$, $\gamma^{(3)}_{E,-p}$ and $N^{(1)}_{E,p}$ at the horizon. When the inner product of the last term of (\ref{psi1horizon}) is computed at the horizon by using (\ref{D9})-(\ref{D13}), we obtain at $\eta=2/\sqrt{a}$ 
\begin{equation}
(\psi^{(1)}_{E,-p},\psi^{(1)}_{E', -p'})_{\text{H}}=16\pi^3 a \, \delta(E-E')\delta(p-p'). 
\end{equation}
Here H stands for horizon. 
It can be shown that the other  inner products, $(\psi^{(2)}_{E,p},\psi^{(2)}_{E', p'})_{\text{H}}$, $(\psi^{(3)}_{E,-p},\psi^{(3)}_{E', -p'})_{\text{H}}$ and $(\psi^{(4)}_{E,p},\psi^{(4)}_{E', p'})_{\text{H}}$ have the same forms as above at the horizon. Thus, although inner products of $\Phi^{\text{II}(\pm \nu)}_{E,p}$ and $\tilde{\Phi}^{\text{II}(\pm \nu)}_{E,p}$ are discontinuous at the horizon $\eta=2/\sqrt{a}$,  the inner products of `the special linear combinations $\psi^{(i)}_{E,p}$' are continuous at the horizon. This ensures that the quantum state is smooth in the vicinity of the horizon for the solution of the matching condition. 

On the other hand $\psi^{(i)}_{E,p}$'s depend on $\gamma^{(i)}_{E,p}$'s away from the horizon, because equations which relate $\Phi^{\text{II}(\nu)}_{E,p}$ and $\tilde{\Phi}^{\text{II}(\nu)}_{E,p}$ are not valid there. Instead (\ref{dep1}) and (\ref{dep2}) hold.  By using (\ref{g13}),  (\ref{g13s}), (\ref{gamma12}) and (\ref{gamma14}) the following expression is obtained away from the horizon.
\begin{equation}
\psi^{(1)}_{E,-p}= N^{(1)}_{E,p} \, [ \sqrt{F^{(1)}_1(E,p)}e^{i\delta_1} \Phi^{\text{II}(\nu)}_{E,-p}+
\sqrt{F^{(1)}_2(E,p)}e^{i\delta_2} \tilde{\Phi}^{\text{II}(\nu)}_{E,-p}] \label{ps1}
\end{equation}
This has the correct K-G norm (\ref{wouldbe}) owing to (\ref{FF}), (\ref{table1}) and (\ref{table4}). Similarly, the other normal modes are found to be
\begin{eqnarray}
\psi^{(2)}_{E,p}&= &N^{(2)}_{E,p} \, [ \sqrt{F^{(2)}_1(E,p)}e^{i\delta_3} \tilde{\Phi}^{\text{II}(\nu)}_{E,p}+
\sqrt{F^{(2)}_2(E,p)}e^{i\delta_4} \Phi^{\text{II}(\nu)}_{E,p}], \label{ps2}\\
\psi^{(3)}_{E,-p}&= &N^{(3)}_{E,p} \, [ \sqrt{F^{(1)}_1(E,p)}e^{i\delta_1} \tilde{\Phi}^{\text{II}(\nu)}_{-E,p}+
\sqrt{F^{(1)}_2(E,p)}e^{i\delta_2} \Phi^{\text{II}(\nu)}_{-E,p}], \label{ps3} \\
\psi^{(4)}_{E,p}&= &N^{(4)}_{E,p} \, [ \sqrt{F^{(2)}_1(E,p)}e^{i\delta_3} \Phi^{\text{II}(\nu)}_{-E,-p}+
\sqrt{F^{(2)}_2(E,p)}e^{i\delta_4} \tilde{\Phi}^{\text{II}(\nu)}_{-E,-p}]. \label{ps4}
\end{eqnarray}
These also have correct K-G norm (\ref{wouldbe}). 
This means that the matching condition cannot determine the quantum scalar field inside the horizon uniquely.  
The near-horizon normal mode, the last term of  (\ref{psi1horizon}), which is unique, is connected to  the deep-inside mode (\ref{ps1}). The latter is not unique and depends on arbitrary functions $F^{(1)}_n(E,p)$ $(n=1,2)$.  Because there is no isometry for  the time variable $\eta$ inside the horizon it is not possible to distinguish between positive- and negative-frequency solutions inside the horizon. So the number of independent normal modes is doubled and as a result the K-G inner product (\ref{norm1}) contains a term with a negative sign. Therefore solutions which contain continuous parameters can exist. 
This shows that the matching condition\cite{Raju1} is not sufficient for determining the bulk inside the horizon. It is necessary to impose yet additional condition to carry out bulk reconstruction. In the next section a natural choice of these parameters will be proposed. 

Because the mode functions $\psi^{(i)}_{E,p}$ ($i=1,2,3,4$) are defined for either $p \geq 0$ or $p \leq 0$, these must satisfy continuity conditions at $p=0$. 
\begin{eqnarray}
\psi^{(1)}_{E,p=0}&=&\psi^{(2)}_{E,p=0}, \label{cont1} \\
\psi^{(3)}_{E,p=0}&=&\psi^{(4)}_{E,p=0}  \label{cont2}
\end{eqnarray}
It can be shown that if $W^{(i)}(E,p=0) \neq 0$ $(i=1,2$), where $W^{(i)}$ is defined in (\ref{W1}) and (\ref{W2}), the following relations must be satisfied.
\begin{eqnarray}
&e^{i(\delta_1(E,p=0)-\delta_2(E,p=0))} = e^{-i(\delta_3(E,p=0)-\delta_4(E,p=0))}, \\
& \frac{N^{(1)}_{E,p=0}}{N^{(3)}_{E,p=0}} = \frac{N^{(2)}_{E,p=0}}{N^{(4)}_{E,p=0}}
\end{eqnarray}
On the contrary it can be shown that  at the horizon the above conditions (\ref{cont1}) and (\ref{cont2}) are satisfied without constraints. 

\section{Determination of $F^{(1)}_n$, $F^{(2)}_n$ and the Propagator for Two Points on Both Sides of the Horizon}
\hspace*{5mm}
As was seen in the previous section only the matching condition at the horizon does not determine the normal modes behind the horizon uniquely. If this ambiguity cannot be removed,  holographic correspondence of the interior to the boundary CFT's is not realized. It is  necessary to impose additional conditions. Here a set of conditions for removing the ambiguity is proposed, and by using the mode expansion of the scalar field inside the horizon, a scalar field propagator with one point inside the horizon and the other outside will be derived in an integral form. 

Let us consider the normal mode (\ref{ps1}) away from the horizon. From the asymptotic formulas (\ref{limitPII}) and   (\ref{limitPIItilde}) for $\Phi^{\text{II}(\nu)}_{E,p}$ and $\tilde{\Phi}^{\text{II}(\nu)}_{E,p}$, respectively,  it is noticed that the first term of (\ref{ps1}) behaves as $\eta^{-ip/\sqrt{a}}$ and the second one as $\eta^{+ip/\sqrt{a}}$. Each terms of (\ref{ps2}), (\ref{ps3}) and (\ref{ps4}) also behave in the same way. In these four $\psi$'s $p$ is a positive number, and the term $\eta^{-ip/\sqrt{a}}=e^{-i(p/\sqrt{a})\ln \eta}$ may be regarded as a `positive frequency' term in the large $\eta$ asymptotic limit. This is a boundary condition at the future. 
So if we set $F^{(1)}_{2}(E,p)=F^{(2)}_{2}(E,p)=0$, all normal modes (\ref{ps1})- (\ref{ps4}) are  `positive frequency' modes. Then 
$F^{(1)}_1(E,p)=F^{(2)}_1(E,p)=e^{\beta p}-1$ is obtained due to (\ref{FF}). For simplicity by setting $N^{(i)}_{E,p}=1$ ($i=1, \dots,4$) and $\delta_n(E,p)=0$ ($n=1,\dots,4$), we have
\begin{eqnarray}
\psi^{(1)}_{E,-p}&= & \sqrt{e^{\beta p}-1}\, \Phi^{\text{II}(\nu)}_{E,-p}, \label{pps1}\\
\psi^{(2)}_{E,p}&= & \sqrt{e^{\beta p}-1} \, \tilde{\Phi}^{\text{II}(\nu)}_{E,p}, \label{pps2}\\
\psi^{(3)}_{E,-p}&= & \sqrt{e^{\beta p}-1} \tilde{\Phi}^{\text{II}(\nu)}_{-E,p}, \label{pps3} \\
\psi^{(4)}_{E,p}&= & \sqrt{e^{\beta p}-1} \, \Phi^{\text{II}(\nu)}_{-E,-p}. \label{pps4}
\end{eqnarray}
This is the simplest set of normal modes. Then the mode expansion (\ref{PhiII2}) contains only these `positive frequency' normal modes and their complex conjugates. We propose to adopt the above conditions for $F$'s and remove the ambiguity as (\ref{pps1})-(\ref{pps4}).

It is also possible to show that the ambiguity associated with $F^{(i)}_{n}(E,p)$ is related to Bogoliubov transformations.  Under the Bogoliubov transformations operators $c_R(E,p)$ and $c_L(E,p)$ and their hermitian conjugates (h.c.) are transformed into new operators, $d_R(E,p)$ and $d_L(E,p)$ and their h.c.. By adjusting suitable transformations the mode expansion for the scalar field (\ref{PhiII2}) is transformed to 
\begin{eqnarray}
&&\Phi^{\text{II}}(\eta,t,x) \nonumber \\
&&= \frac{1}{4\pi \sqrt{\pi a}} \, \int_0^{\infty}dE\int_0^{\infty}dp \, \sqrt{e^{\beta p}-1}
\Big[ d_R(E,-p) \Phi^{\text{II}(\nu)}_{E,-p}+d_R(E,p) \tilde{\Phi}^{\text{II}(\nu)}_{E,-p}
 \nonumber \\
&& +d_L(E,p) \Phi^{\text{II}(\nu)}_{-E,-p}+d_L(E,-p) \tilde{\Phi}^{\text{II}(\nu)}_{-E,p}+h.c.\Big]. \label{ddphi}
\end{eqnarray} 
The above expansion is a result of four sets of transformations. For example, 
\begin{multline}
d_R(E,-p)= (e^{\beta p}-1)^{-1/2} \big[ c_R(E,-p)N^{(1)}\sqrt{F^{(1)}_1}e^{i\delta_1} \\+c_L^{\dagger}(E,-p)N^{(3)\ast}\sqrt{F^{(1)}_2}e^{\beta p/2}e^{-i\delta_2}e^{-i\pi(\nu+1)}\big],  
\end{multline}
and $d_R^{\dagger} (E,-p)$ is given by the hermitian conjugate of the above equation. Now $F^{(i)}_n$'s have disappeared in (\ref{ddphi}). The ambiguity of the solution which satisfies the matching condition coincides with the parameters of these transformations. 
New operators, $d_R$ and $d_L$, however, do not have the thermal averages like (\ref{VEV}) in the state $|\Psi_{TFD}\rangle_{\beta}$, 
unless\footnote{Even if eqs (\ref{FFcond}) are not satisfied, the mode expansion (\ref{ddphi}) defines a quantum scalar theory at finite temperature inside the horizon, because the scalar field (\ref{PhiII2}) is expressed in terms of $c_{L,R}$'s.} 
\begin{equation}
F^{(1)}_1(E,p)=F^{(2)}_1(E,p)=e^{\beta p}-1, \qquad F^{(1)}_{2}(E,p)=F^{(2)}_{2}(E,p)=0. \label{FFcond}
\end{equation}

With the choice (\ref{pps1})-(\ref{pps4}) a propagator for  two points in regions I and II, respectively, are given by 
\begin{eqnarray}
&& {}_{\beta}\langle  \Psi_{TFD}| \Phi^{\text{I}}(t_1,y_1,x_1) \Phi^{\text{II}}(\eta_2, t_2,x_2)|\Psi_{TFD}\rangle_{\beta} \nonumber \\
&=& \frac{1}{16\pi^3 a}\int_{0}^{\infty}dp \int_0^{\infty} dE \,
\sqrt{
\frac{e^{\beta p}-1}{1-e^{-\beta E}}  } \Big[ \Phi^{\text{I}}_{E,p} \, \tilde{\Phi}^{\text{II}(\nu)\ast}_{E,p}
+\Phi^{\text{I}}_{E,-p} \, \Phi^{\text{II}(\nu)\ast}_{E,-p} 
+e^{-\beta E}\,  \Phi^{\text{I}\ast}_{E,p} \, \tilde{\Phi}^{\text{II}(\nu)}_{E,p} \nonumber \\
&& \qquad +e^{-\beta E}\,  \Phi^{\text{I}\ast}_{E,-p} \, \Phi^{\text{II}(\nu)}_{E,-p} 
 +
e^{-\beta E/2} \, \Phi^{\text{I}}_{E,p} \, \Phi^{\text{II}(\nu)}_{-E,-p}
+e^{-\beta E/2} \, \Phi^{\text{I}}_{E,-p} \, \tilde{\Phi}^{\text{II}(\nu)}_{-E,p} \nonumber \\
&& \qquad +e^{-\beta E/2} \, \Phi^{\text{I}\ast}_{E,p} \, \Phi^{\text{II}(\nu)\ast}_{-E,-p} 
+e^{-\beta E/2} \, \Phi^{\text{I}\ast}_{E,-p} \, \tilde{\Phi}^{\text{II}(\nu)\ast}_{-E,p} 
\Big]  \label{propa}
\end{eqnarray}
This is obtained by using (\ref{PhiR4}), (\ref{PhiII2}) and (\ref{VEV}). 
An explicit calculation of the integral on the right-hand side is not carried out yet. It is, however, possible to give an argument that this propagator will not be the same function as those obtained in \cite{Ichinose} and \cite{GT}. 
Actually, if the propagator is a function only of the geodesic distance, it  will take the form 
\begin{eqnarray}
G&=&\frac{1}{16\pi^3 a}\int_{-\infty}^{\infty}dp \int_0^{\infty} dE\Big[ \Phi^{\text{I}}_{E,p} \, \Phi^{\text{II}(\nu)\ast}_{E,p} +\Phi^{\text{I}}_{-E,-p} \, \Phi^{\text{II}(\nu)\ast}_{-E,-p}\Big]
\nonumber \\
&& = \frac{1}{16\pi^3 a}\int_{-\infty}^{\infty}dp \int_0^{\infty} dE\Big[ \Phi^{\text{I}}_{E,p} \, \Phi^{\text{II}(\nu)\ast}_{E,p} +e^{-\pi i(\nu+1)} e^{-\beta( E+p)/2}\Phi^{\text{I}\ast}_{Ep} \, \Phi^{\text{II}(\nu)}_{E,p}\Big]. \label{AdSp}
\end{eqnarray}
This is because (\ref{AdSp}) can be rewritten as
\begin{equation}
G=\int_0^{\infty}\frac{d\omega}{4\pi}\int_{|k|\leq \omega}dk \, \Phi^{\text{I}}_{\omega,k}(t_1,y_1,x_1) \Phi^{\text{II}(\nu)\ast}_{\omega,k}(\eta_2,t_2,x_2), \label{Gomegak}
\end{equation}
where $\Phi^{\text{I}}_{\omega,k}$ and $\Phi^{\text{II}(\nu)}_{\omega,k}$ are defined in (\ref{phiI}) and (\ref{phiII}), respectively,  
and in tern this is computed as a correlation function $\langle 0| \phi(\tau_1,u_1,\chi_1)\phi
(\tau_2,u_2,\chi_2)|0\rangle$ in  AdS$_3$ spacetime\footnote{For notation see Appendix A.} followed by coordinate transformations (\ref{I}), (\ref{II}), and also  by analytic continuation $t_2 \rightarrow t_2-i\beta/4$. 
The result is \cite{Kaplan}, \cite{Ichinose}
\begin{equation}
G=(1/2\pi) \rho^{(\nu+1)/2}/(1-\rho), \label{Geq}
\end{equation}
where
\begin{eqnarray}
\rho&=&\frac{1}{\xi^2}[1-\sqrt{1-\xi^2}]^2\equiv e^{-2\sigma}, \\
\xi &=& \frac{4\sqrt{a} y_1(a\eta^2_2+4)}{4\sqrt{a}\eta_2(4+ay_1^2)\cosh (\sqrt{a}(x_{1}-x_2))   -(4-ay_1^2)(a\eta_2^2-4)\sinh (\sqrt{a}(t_{1}-t_2))}. \label{geod}
\end{eqnarray}
Here $\sigma$ is a geodesic distance between the two points. 

Because the normal modes in region II, $\psi^{(i)}_{E,p}$, contains $\tilde{\Phi}^{\text{II}(\nu)}_{E,p}$ in addition to $\Phi^{\text{II}(\nu)}_{E,p}$, the propagator (\ref{propa}) inevitably contains this additional mode, and the propagator cannot be transformed into the integral form (\ref{Gomegak}).\footnote{See the footnote 2 for the $(\omega,k)$ representation of $\tilde{\Phi}^{\text{II}(\nu)}_{E,p}$.} 
At the very least, this propagator will not be obtained from that in AdS$_3$ by coordinate transformations.  Even for  choices of $F^{(1)}_{1,2}$ and $F^{(2)}_{1,2}$ other than (\ref{FFcond}), this observation will be also valid. If the integral for the propagator could be computed explicitly, this claim would be established. This will be left for study in the future. 
It is also interesting to study the propagator with both points inside the horizon.  

Finally, the two-point function between fields on both sides of the horizon in BTZ black hole background which is periodic for $\varphi \rightarrow \varphi +2\pi$ is obtained by replacing $x$ by $\varphi$ and summing over images of (\ref{propa}) obtained by a shift of $\varphi_1$. 
\begin{equation}
G^{\text{I},\text{II}}(t_1,y_1,\varphi_1; \eta_2, t_2,\varphi_2)= \sum_{n=-\infty}^{\infty}   {}_{\beta}\langle \Psi_{ TFD}|\Phi^{\text{I}}(t_1,y_1,\varphi_1+2\pi n) \Phi^{\text{II}}(\eta_2, t_2,\varphi_2)|\Psi_{TFD}\rangle_{\beta}
\end{equation}

\section{Summary and Discussions}
\hspace{5mm}
In this section the results of this paper will be briefly summarized and some discussions will be given. 

In this paper quantization of a free scalar field in BTZ black hole background including the region inside the horizon is studied. The  normal modes of a scalar field  in black hole background are obtained from those of AdS$_3$ spacetime by suitable coordinate transformations. These normal modes turn out to be not eigenstates of energy and momentum. We quantized a scalar field on the same constant-$t$ slice $\Sigma$ in both regions I and III: K-G inner products are computed on the constant $t$ slice $\Sigma$ obtained by combining those in both regions. By changing the basis of the normal modes to that of eigenstates of energy and momentum it is found that the creation and annihilation operators  in each regions I and III can be identified as CFT primary operators in boundary CFTs. The vacuum state is shown to be the TFD (\ref{TFD}). 

Then a  scalar field behind a horizon of a two-sided BTZ black hole is quantized by using the matching condition of \cite{Raju1} for normal modes. It is found that the scalar field just inside the horizon is expanded in terms of the normal modes, $\Phi^{\text{II}(\nu)}_{E,p}$ and $\Phi^{\text{II}(-\nu)}_{E,p}$, while deep inside the horizon it is expanded into $ \Phi^{\text{II}(\nu)}_{E,p}$ and $\tilde{\Phi}^{\text{II}(\nu)}_{E,p}$. 
The result that there exist  two independent sets of normal modes inside the horizon is in accord with the fact that particles come into region II from both regions I and III.
It is shown that the scalar field can be connected across the horizon by using the matching condition and the scalar field also satisfies correct equal time commutation relations behind the horizon.  It is, however, shown that there are still undetermined coefficients in the normal modes for a scalar field deep inside the horizon. The matching condition at the horizon is not sufficient to determine a scalar theory  inside the horizon of BTZ black hole uniquely. It is found that a mode expansion of a scalar field  behind the horizon depends on extra functions $F^{(i)}_n(E,p)$. Then we imposed a certain appropriate condition (\ref{FFcond}) on $F^{(i)}_n(E,p)$ and obtained a new propagator between two points on opposite sides of the horizon, respectively. This is a boundary condition at the future. 
The boundary limit of this propagator will not coincide with the propagator of \cite{Ichinose}, \cite{GT} and \cite{KOS} , because the mode functions (\ref{ps1})-(\ref{ps4}) are composed of both $\Phi_{E,p}^{\text{II}(\nu)}$ and $\tilde{\Phi}_{E,p}^{\text{II}(\nu)}$.  If the resolution of the ambiguity of the internal modes in sec.9 is appropriate, then the boundary CFT's have the information on the bulk operators inside. It will be possible to obtain information on the structure of the interior of the horizon by studying the propagator (\ref{propa}). On the other hand, if the propagator (\ref{propa}) does not coincide with those of \cite{Ichinose}, \cite{GT} and \cite{KOS}, however, then this may imply that the propagator (\ref{propa}) might have some non-analyticity at the horizon. Then it may be necessary to examine the validity of the matching condition and to seek its modification. 

 In this paper quantization of a scalar field in region IV (past) is not discussed. This can be also  carried out without difficulty in a similar way to the procedure adopted in region II. In this case  the lower sign of (\ref{eta2}) needs to be used for a time variable $\eta$. Then $\eta$ takes values in $0 <\eta \leq 2/\sqrt{a}$. Then, the coordinate transformations (\ref{IV1})-(\ref{IV3}) will give a normal mode $\Phi^{\text{IV}}_{\omega,k}(\eta,t,x)$ and  the solution to the matching condition can be studied. No new restrictions to the theory itself except for results similar to those in region II  are expected from the matching condition for regions IV and outside. 

In \cite{GT}  bulk local states for a scalar field behind and outside the  BTZ black hole were constructed and the operators corresponding to these bulk local states were shown to be dual to the CFT primary operators on the boundary of the past half torus for the Euclidean path integral defining a TFD state. This approach differs from the one of this paper where the operators in the black hole interior are supposed to be dual to those on the timelike cylindrical boundary at the spatial infinity. 
It may be possible to get information of the scalar field operator from the bulk local scalar state of \cite{GT}.  
It is interesting to obtain the normal mode expansion of the scalar field explicitly and check the canonical commutation relations of the field behind the horizon.  

The result of this paper will have important implication on quantization of matter fields in higher-dimensional AdS black hole backgrounds. It is interesting to study the structure of matter theories behind the horizon in higher-dimensional black holes. As was found for the BTZ black hole in this paper, normal modes of a scalar field inside and near the horizon of higher-dimensional black holes may be distinct from those deep inside the horizon. The normal modes deep inside the horizon may have interesting properties. Because there are severe curvature singularities behind the horizon of higher-dimensional black holes, reliable analysis may be possible only in  the region of small curvature and the complete analysis may not be easy. However, the normal modes inside the horizon away from the singularity may be studied by some approximation methods. Then in higher dimensions it will be necessary to use the matching condition at the horizon for quantization of a scalar field. So it might be necessary to cope with the ambiguity of the normal mode expansion like that found in this paper.  In such a case it would be necessary to come up with a new principle to remove the ambiguity of  the scalar normal modes inside the horizon to comply with the holographic principle. Because the singularity is stronger than in 3 dimensions, it is not clear whether it is also appropriate to impose a future boundary condition. 
This issue needs to be studied further.

\setcounter{section}{0}
\renewcommand{\thesection}{\Alph{section}}

\section{Connection of Normal Modes of a Scalar Field in BTZ Black Hole to those in AdS$_3$ }
\hspace{5mm}
The metric for massless BTZ black hole is given by
\begin{equation}
ds^2 = \frac{1}{u^2} \, (du^2+dw^+ dw^-)   \label{uniformization}
\end{equation}
Here AdS length is set to unity. $u$ is a radial coordinate and $w^+=\chi+\tau$ and $w^-=\chi-\tau$ and $\tau$ is a time, and $\chi$ the spatial one. 
We perform the following coordinate transformations  $(u,w^+,w^-) \rightarrow (y,z^+,z^-)$ in  the above equation.\cite{Roberts}\cite{Banados}\cite{Kaplan}
\begin{eqnarray}
w^+ &=& f(z^+)-\frac{2y^2(f'(z^+))^2\bar{f}''(z^-)}{4f'(z^+)\bar{f}'(z^-)+y^2f''(z^+)\bar{f}''(z^-)}, \label{transw} \\
w^- &=& \bar{f}(z^-)-\frac{2y^2(\bar{f}'(z^-))^2f''(z^+)}{4f'(z^+)\bar{f}'(z^-)+y^2f''(z^+)\bar{f}''(z^-)}, \label{transwb} \\
u &=& \frac{4y(f'(z^+)\bar{f}'(z^-))^{3/2}}{4f'(z^+)\bar{f}'(z^-)+y^2f''(z^+)\bar{f}''(z^-)}
\label{transu}
\end{eqnarray}
Then we obtain a new metric. 
\begin{eqnarray}
ds^2 =&& \frac{1}{y^2}(dy^2+dz^+dz^-)-\frac{1}{2}S[f,z^+](dz^+)^2-\frac{1}{2}S[\bar{f},z^-](dz^-)^2 \nonumber \\&&+y^2\frac{1}{4}S[f,z^+]S[\bar{f},z^-]dz^+dz^- \label{metricS}
\end{eqnarray}
Here $S[f,z]$ is a Schwarzian derivative.
\begin{equation}
S[f,z]=\frac{f'''(z)}{f'(z)}-\frac{3}{2}\Big(\frac{f''(z)}{f'(z)}\Big)^2
\end{equation}

Let us next consider the transformations generated by $f(z^+)=e^{\sqrt{a}z^+}$ and $\bar{f}(z^-)=e^{\sqrt{a}z^-}$.
\begin{eqnarray}
w^+ &=& \frac{4-ay^2}{4+ay^2}\ e^{\sqrt{a}z^+}, \nonumber \\
w^- &=& \frac{4-ay^2}{4+ay^2}\ e^{\sqrt{a}z^-}, \nonumber\\
u &=& \frac{4\sqrt{a}y}{4+ay^2} e^{\frac{1}{2}\sqrt{a}(z^++z^-)} \label{I}
\end{eqnarray}
Here $a=4GM$ is a parameter related to the black hole mass $M$, and  $G$ is a 3d Newton constant. 
Then the metric is transformed to 
\begin{eqnarray}
ds^2 &=&   \frac{1}{y^2}dy^2+\frac{1}{4}(a(dz^+)^2+a(dz^-)^2)+(\frac{1}{y^2}+\frac{1}{16}a^2y^2)dz^+dz^-  \label{BTZout}
\end{eqnarray}
When we define $z^+=x+t$, $z^-=x-t$ and 
\begin{equation}
r \equiv \frac{1}{y}+\frac{a}{4}y, \label{ry} 
\end{equation}
then the metric reads
\begin{equation}
ds^2= \frac{1}{r^2-a}dr^2-(r^2-a)dt^2+r^2dx^2 \label{Schwarzschild}
\end{equation}
Because $r \geq \sqrt{a}$, 
(\ref{BTZout}) describes the space-time outside the black hole.  (\ref{ry}) is solved for $y$ as
\begin{equation}
y = \frac{2}{a} (r \pm \sqrt{r^2-a})
\end{equation}
The horizon is located at $r=r_+=\sqrt{a}$. 
Each $0 < y \leq \frac{2}{\sqrt{a}}$ and $\frac{2}{\sqrt{a}} \leq y <\infty$ correspond to the exterior region $ r \geq \sqrt{a}$.

Maximally extended Schwarzschild space time consists of four regions: right (I) region and left (III) one, which are spacetimes outside the horizon. Future (II) and past (IV) one, which are behind the horizon. The normal modes of a scalar field in BTZ background are obtained from those in pure AdS$_3$ space (\ref{uniformization})  by coordinate transformations. 

A classical equation of motion for a real scalar field $\phi$ with mass $m$ in (\ref{uniformization}) is given by 
\begin{equation}
\partial_u^2 \phi-\frac{1}{u}\partial_u \phi+(\partial_{\chi}^2-\partial_{\tau}^2)\phi-\frac{m^2}{u^2}\phi=0. \label{KGeqAdS}
\end{equation}
By separation of variables mode functions of this scalar field will be obtained in the form 
\begin{equation}
\phi_{\omega k}(\tau,u,\chi) =e^{-i\omega \tau+ik\chi} \, f_{\omega k}(u),
\end{equation}
where $\omega$ and $k$ are constants, and $f_{\omega k}(u)$ is a solution to the following equation
\begin{equation}
f''_{\omega k}(u)-\frac{1}{u}f'_{\omega k}(u)+\Big(\omega^2-k^2-\frac{m^2}{u^2}\Big)f_{\omega k}(u)=0
\end{equation}
By BDHM dictionary\cite{BDHM} the scalar field dual to the 2d CFT must satisfy a boundary condition $\phi \sim u^{1+\nu}$ near the boundary $u \rightarrow 0$, where $\nu=\sqrt{1+m^2}$, and for non-integer $\nu$, this determines $f_{\omega k}(u)=uJ_{\nu}(\sqrt{\omega^2-k^2} \, u)$. The mode functions are then given by 
\begin{equation}
\phi_{\omega k}(\tau,u,\chi) =e^{-i\omega \tau+ik\chi} \, uJ_{\nu}(\sqrt{\omega^2-k^2} \, u), \quad (\omega \geq 0, \ -\infty < k< \infty ). \label{AdSModes}
\end{equation}
Here $J_{\nu}(z)$ is a Bessel function of the $\nu$-th order. 

In region (I) the map connecting the uniformization coordinates and those of black hole space-time is given by (\ref{I}). In this region $y$ takes values in $0<y \leq \frac{2}{\sqrt{a}}$. Hence the mode functions in region I are 
\begin{equation}
\Phi^{\text{I}}_{\omega,k}(t,y,x) \equiv  \frac{4\sqrt{a}y}{4+ay^2} e^{\sqrt{a}x} J_{\nu}\Big(\mu\frac{4\sqrt{a}y}{4+ay^2}e^{\sqrt{a}x}\Big) \exp \big[ i \frac{4-ay^2}{4+ay^2}e^{\sqrt{a}x} (k\cosh \sqrt{a}t-\omega \sinh \sqrt{a}t)\big]
\label{phiI2}
\end{equation}
Here $\mu=\sqrt{\omega^2-k^2}$. 

In region (III) we need to make a shift $t \rightarrow t-i\beta/2$ in (\ref{I}). 
However, $y$ must be mapped to the region $\frac{2\pi}{\sqrt{a}} \leq y$ by a transformation $y \rightarrow \frac{4}{ay}$. Then the relation (\ref{I}) is unchanged. To obtain the normal modes in region III the direction of time must be flipped:$t \rightarrow -t$ in (\ref{I}). 
Then the normal modes in region III are given by
\begin{multline}
\Phi^{\text{III}}_{\omega,k}(t,y,x) 
= \frac{4\sqrt{a}y}{4+ay^2} e^{\sqrt{a}x} J_{\nu}\Big(\mu\frac{4\sqrt{a}y}{4+ay^2}e^{\sqrt{a}x}\Big) \\
\exp \big[ i \frac{4-ay^2}{4+ay^2}e^{\sqrt{a}x }(k\cosh \sqrt{a}t+\omega \sinh \sqrt{a}t)\} \big]
\label{phiIII2}
\end{multline}

To describe the interior of the horizon, we introduce a new radial coordinate $\eta (>0)$ by 
\begin{eqnarray}
y &=& \frac{8\eta}{a\eta^2+4}+i\frac{2(a\eta^2-4)}{\sqrt{a}(a\eta^2+4)} \label{yeta}
\end{eqnarray}
Note that this is a complex transformation. Then the metric is transformed to 
\begin{eqnarray}
ds^2 &=& \frac{a(a\eta^2-4)^2}{(a\eta^2+4)^2} dt^2-\frac{16a}{(a\eta^2+4)^2}d\eta^2+\frac{16 a^2\eta^2}{(a\eta^2+4)^2}dx^2 \label{dseta}
\end{eqnarray}
Now $t$ is a space-like variable and $x$ the time-like one. Note that $\eta=0$ and $\eta=\infty$ are singularities. This metric describes the region behind the horizon of the spacetime  (\ref{Schwarzschild}), This  can also be confirmed by
\begin{eqnarray}
r^2-a &=& -a \Big( \frac{a\eta^2-4}{a\eta^2+4}\Big)^2 \leq 0
\end{eqnarray}
The relation between $\eta$ and $r$ is also two-fold,
\begin{equation}
\eta = \frac{2}{r}\Big(1 \pm \frac{1}{\sqrt{a}}\sqrt{a-r^2}\Big) \label{eta2}
\end{equation}

To go to region (II) the transformation (\ref{yeta}) is used. 
The relation between $\eta$ and $r$ is given by
\begin{equation}
\eta = \frac{2}{r}\Big(1 + \frac{1}{\sqrt{a}}\sqrt{a-r^2}\Big) \label{eta2II}
\end{equation}
The range of $\eta$ is $2/\sqrt{a}< \eta < \infty$. In addition the shift $t \rightarrow t-i\beta/4$ must be carried out in order to make coordinates real-valued. Here $\beta$ is the inverse temperature. Hence we have  
\begin{eqnarray}
u &=& \frac{1}{4 \sqrt{a}}(a\eta+\frac{4}{\eta}) \, e^{\frac{1}{2}\sqrt{a}(z^++z^-)}, \nonumber \\
w^+ &=& -i \frac{4-ay^2}{4+ay^2} e^{\sqrt{a}z^+}=-\frac{a\eta^2-4}{4 \sqrt{a}\eta} e^{\sqrt{a}z^+}, \nonumber \\
w^- &=& i \frac{4-ay^2}{4+ay^2} e^{\sqrt{a}z^-}=\frac{a\eta^2-4}{4\sqrt{a}\eta} e^{\sqrt{a}z^-} \label{II}
\end{eqnarray}
Normal modes in region II are given by
\begin{equation}
\Phi^{\text{II}}_{\omega, k}(\eta, t,x)
= \frac{4+a\eta^2}{4\sqrt{a}\eta}\, e^{\sqrt{a}x}J_{\nu}(\mu \frac{4+a\eta^2}{4\sqrt{a}\eta}e^{\sqrt{a}x})\exp\big[ i\frac{a\eta^2-4}{4\sqrt{a}\eta} e^{\sqrt{a}x}(\omega \cosh \sqrt{a}t  -k\sinh \sqrt{a}t)\big]. \label{phiII2}
\end{equation}

Finally, in region (IV) we need to shift $t \rightarrow t-3i\beta/4
$ in the results (\ref{II}). 
\begin{eqnarray}
u &=& \frac{1}{4 \sqrt{a}}(a\eta+\frac{4}{\eta}) \, e^{\frac{1}{2}\sqrt{a}(z^++z^+)}, \label{IV1} \\
w^+ &=& i \frac{4-ay^2}{4+ay^2} e^{\sqrt{a}z^+}=\frac{a\eta^2-4}{4 \sqrt{a}\eta} e^{\sqrt{a}z^+}, \\
w^- &=& -i \frac{4-ay^2}{4+ay^2} e^{\sqrt{a}z^-}=-\frac{a\eta^2-4}{4 \sqrt{a}\eta} e^{\sqrt{a}z^-} \label{IV3}
\end{eqnarray}
Region for $\eta$ is  $0 \leq \eta \leq 2/\sqrt{a}$.

\section{Functions $g(E,p)$ and $\tilde{g}(E,p)$}
\hspace*{5mm}
In this appendix some formulas related to the functions $g(E,p)$ and $\tilde{g}(E,p)$ are presented. 
$g(E,p)$ is defined in (\ref{gEp}). By using a representation of the modified Bessel function of the second kind, 
\begin{equation}
K_{\nu}(x)=\frac{1}{2} \, e^{\nu \pi i/2} \int_{-\infty}^{\infty}e^{-ix \sinh t+\nu t}dt,
\end{equation}
this function  is evaluated as
\begin{equation}
g(E,p)= 2^{1-\nu}a^{\Delta/2} \, [\Gamma(\Delta)]^{-1} \, e^{\frac{\pi}{2\sqrt{a}} \, E} \, \int_0^{\infty} \, \mu^{\nu-\frac{i}{\sqrt{a}}p} \, K_{\frac{iE}{\sqrt{a}}}(\mu)d\mu.  \label{gfunc}
\end{equation}
Furthermore by using a formula
\begin{equation}
\int_0^{\infty} x^{\mu-1} \, K_{\nu}(ax)dx= 2^{\mu-2}a^{-\mu} \, \Gamma\big(\frac{\mu-\nu}{2}\big) 
\, \Gamma\big(\frac{\mu+\nu}{2}\big), 
\end{equation}
which is valid for $\text{Re} \mu> |\text{Re} \nu|$, the following equation is finally obtained. 
\begin{equation}
g(E,p)= \frac{a^{\Delta/2}}{\Gamma(\Delta)} e^{\frac{\pi}{2\sqrt{a}}E} \, 2^{-\frac{ip}{\sqrt{a}}} \, \Gamma\big( \frac{1}{2}(\Delta-i\frac{p}{\sqrt{a}}-i\frac{E}{\sqrt{a}})\big)
\Gamma\big( \frac{1}{2}(\Delta-i\frac{p}{\sqrt{a}}+i\frac{E}{\sqrt{a}})\big)
\end{equation}
Then the following quantity related to the normalization factor is positive semi-definite. 
\begin{multline}
|g(E,p)|^2-|g(-E,-p)|^2  \\
= \frac{a^{\Delta}}{(\Gamma(\Delta))^2} \, \big(e^{(\beta/2) E}-e^{-(\beta/2) E}\big) \, \big| \Gamma\big( \frac{1}{2}(\Delta-i\frac{p}{\sqrt{a}}-i\frac{E}{\sqrt{a}})\big) \big|^2 \, \big| \Gamma\big( \frac{1}{2}(\Delta-i\frac{p}{\sqrt{a}}+i\frac{E}{\sqrt{a}})\big) \big|^2
\end{multline}

In a similar fashion the following results can be derived for the function $\tilde{g}(E,p)$ (\ref{tildegEp}). 
\begin{equation}
\tilde{g}(E,p)= \frac{a^{\Delta/2}}{\Gamma(\Delta)} e^{-\frac{\pi}{2\sqrt{a}}E} \, 2^{-\frac{ip}{\sqrt{a}}} \, \Gamma\big( \frac{1}{2}(\Delta-i\frac{p}{\sqrt{a}}-i\frac{E}{\sqrt{a}})\big)
\Gamma\big( \frac{1}{2}(\Delta-i\frac{p}{\sqrt{a}}+i\frac{E}{\sqrt{a}})\big), 
\end{equation}
\begin{multline}
|\tilde{g}(-E,-p)|^2-|\tilde{g}(E,p)|^2  \\
= \frac{a^{\Delta}}{(\Gamma(\Delta))^2} \, \big(e^{(\beta/2) E}-e^{-(\beta/2) E}\big) \, \big| \Gamma\big( \frac{1}{2}(\Delta-i\frac{p}{\sqrt{a}}-i\frac{E}{\sqrt{a}})\big) \big|^2 \, \big| \Gamma\big( \frac{1}{2}(\Delta-i\frac{p}{\sqrt{a}}+i\frac{E}{\sqrt{a}})\big) \big|^2
\end{multline}

\section{Inner Products of $\Phi^{\text{II}(\pm \nu)}_{E,p}$ and $\tilde{\Phi}^{\text{II}(\pm \nu)}_{E,p}$  for $\eta > 2/\sqrt{a}$}
\hspace*{5mm}
In this appendix the result for the inner product (\ref{innerII1}), $(\Phi^{\text{II}(\nu)}_{E,p}, \Phi^{\text{II}(\nu)}_{E',p'})_{\text{II}}$, will be derived, and  inner products of various normal mode functions will be presented. 
For this purpose the asymptotic form of $\Phi^{\text{II} (\nu)}_{E,p} $ as $\eta \rightarrow +\infty$ is necessary. 
In this limit  $\Phi^{\text{II}(\nu)}_{E,p}$ (\ref{f}) behaves after rescaling $\mu \rightarrow 4\mu/(\sqrt{a}\eta)$ as
\begin{equation}
\Phi^{\text{II}(\nu)}_{E,p} \stackrel{\eta \rightarrow +\infty}{\longrightarrow}  e^{-iEt+ipx} \, (\frac{\sqrt{a}}{4}\eta)^{i\frac{p}{\sqrt{a}}}\int_{-\infty}^{\infty}d\zeta e^{-i\frac{E}{\sqrt{a}}\zeta}\int_{0}^{\infty}d\mu\mu^{-i\frac{p}{\sqrt{a}}} J_{\nu}(\mu)e^{i\mu \cosh \zeta}
\end{equation}
Here $\zeta$ integral is carried out by using a formula 
\begin{equation}
\int_{-\infty}^{\infty}d\zeta e^{-\nu\zeta-z\cosh \zeta}d\zeta=2K_{\nu}(z).  \label{Kintegral}
\end{equation}
Then $\mu$ integration is performed for $Re(a\pm ib)>0$, $Re(\nu-\lambda+1)> |Re \mu|$ by \cite{Grad}
\begin{multline}
\int_0^{\infty}x^{-\lambda}K_{\mu}(ax)J_{\nu}(bx)dx \\
= \frac{b^{\nu}\Gamma(\frac{\nu-\lambda+\mu+1}{2})\Gamma(\frac{\nu-\lambda-\mu+1}{2})}{2^{\lambda+1}a^{\nu-\lambda+1}\Gamma(\nu+1)} F(\frac{\nu-\lambda-\mu+1}{2}, \frac{\nu-\lambda+\mu+1}{2};\nu+1;-\frac{b^2}{a^2})  \label{Grad}
\end{multline}
The following result is obtained.
\begin{multline}
\Phi^{\text{II}(\nu)}_{E,p} \stackrel{\eta \rightarrow +\infty}{\longrightarrow} e^{\pi i(\nu+1)/2}e^{-iEt+ipx} \, 2^{-i\frac{p}{\sqrt{a}}} e^{\beta p/4} \frac{\Gamma(\frac{1}{2}(\nu+1+i\frac{E-p}{\sqrt{a}})) 
\Gamma(\frac{1}{2}(\nu+1-i\frac{E+p}{\sqrt{a}})) }{\Gamma(\nu+1)}\\ 
\times F(\frac{1}{2}(\nu+1+i\frac{E-p}{\sqrt{a}}), \frac{1}{2}(\nu+1-i\frac{E+p}{\sqrt{a}});\nu+1;1) (\sqrt{a}\eta/4)^{ip/\sqrt{a}} \\
=e^{\pi i(\nu+1)/2}e^{-iEt+ipx} \, 2^{-i\frac{p}{\sqrt{a}}} e^{\beta p/4} \, \Gamma\big(\frac{ip}{\sqrt{a}}\big) \, \frac{\Gamma(\frac{1}{2}(\nu+1+i\frac{E-p}{\sqrt{a}})) 
\Gamma(\frac{1}{2}(\nu+1-i\frac{E+p}{\sqrt{a}})) }{\Gamma(\frac{1}{2}(\nu+1-i\frac{E-p}{\sqrt{a}})) 
\Gamma(\frac{1}{2}(\nu+1+i\frac{E+p}{\sqrt{a}})) }(\sqrt{a}\eta/4)^{ip/\sqrt{a}} 
 \label{limitPII}
\end{multline}
This asymptotics also shows that $\Phi^{\text{II}(\nu)}_{E,p}  $ satisfies the following complex conjugation rule.
\begin{equation}
(\Phi^{\text{II}(\nu)}_{E,p} )^{\ast} = e^{-\pi i (\nu+1)} \, e^{\beta p/2} \, \Phi^{\text{II}(\nu)}_{-E,-p}   \label{complexJ}
\end{equation}
This shows that $\Phi^{\text{II}(\nu)}_{E,p}$ does not form a complete set of orthonormal basis of functions in region II. 

Because the inner product does not depend on $\eta$ as far as $\eta >2/\sqrt{a}$, it can be evaluated in the limit $\eta \rightarrow +\infty$ by using (\ref{limitPII}). The following results are obtained.
\begin{equation}
 (\Phi^{\text{II}(\nu)}_{E,p}, \Phi^{\text{II}(\nu)}_{E',p'})_{\text{II}}=\frac{-8\pi^3a}{\sinh (\beta p/2)}e^{\beta p/2}\delta(E-E')\delta(p-p'),
\end{equation}

Similarly, formulas for $\tilde{\Phi}^{\text{II}(\nu)}_{E,p}$ (\ref{ftilde}) can be derived. 
Asymptotic form for (\ref{ftilde}) in the limit $\eta \rightarrow \infty$ is given by
\begin{multline}
\tilde{\Phi}^{\text{II}(\nu)}_{E,p} \stackrel{\eta \rightarrow +\infty}{\longrightarrow} \\
e^{\pi i(\nu+1)/2}e^{-iEt+ipx} \, 2^{i\frac{p}{\sqrt{a}}} e^{-\beta p/4} \, \Gamma\big(\frac{-ip}{\sqrt{a}}\big) \, \frac{\Gamma(\frac{1}{2}(\nu+1+i\frac{E+p}{\sqrt{a}})) 
\Gamma(\frac{1}{2}(\nu+1-i\frac{E-p}{\sqrt{a}})) }{\Gamma(\frac{1}{2}(\nu+1-i\frac{E+p}{\sqrt{a}})) 
\Gamma(\frac{1}{2}(\nu+1+i\frac{E-p}{\sqrt{a}})) }(\sqrt{a}\eta/4)^{-ip/\sqrt{a}} 
 \label{limitPIItilde}
\end{multline}
Inner products of (\ref{ftilde}) are given by 
\begin{equation}
 (\tilde{\Phi}^{\text{II}(\nu)}_{E,p}, \tilde{\Phi}^{\text{II}(\nu)}_{E',p'})_{\text{II}} =\frac{8\pi^3a}{\sinh (\beta p/2)}e^{-\beta p/2}\delta(E-E')\delta(p-p'),
\end{equation}

In the remainder of this appendix, 
inner products of $\Phi^{\text{II}(\pm \nu)}_{E,p}$ and $\tilde{\Phi}^{\text{II}(\pm \nu)}_{E,p}$ are presented. 
\begin{eqnarray}
(\Phi^{\text{II}(\nu)}_{E,p}, \Phi^{\text{II}(\nu)}_{E',p'})_{\text{II}}&=&(\Phi^{\text{II}(-\nu)}_{E,p}, \Phi^{\text{II}(-\nu)}_{E',p'})_{\text{II}}=-8\pi^3a\frac{e^{\beta p/2}}{\sinh \beta p/2} \delta(E-E')\delta(p-p'), \label{table1}\\
(\Phi^{\text{II}(\nu)}_{E,p}, \Phi^{\text{II}(-\nu)}_{E',p'})_{\text{II}}&=&-8\pi^3a\frac{e^{\beta p/2}}{\sinh \beta p/2} e^{-\pi i \nu} D(E,p)\delta(E-E')\delta(p-p'),  \\
(\Phi^{\text{II}(-\nu)}_{E,p}, \Phi^{\text{II}(\nu)}_{E',p'})_{\text{II}}&=&-8\pi^3a\frac{e^{\beta p/2}}{\sinh \beta p/2} e^{\pi i \nu} D^{\ast}(E,p)\delta(E-E')\delta(p-p'), 
\end{eqnarray}
\begin{eqnarray}
(\tilde{\Phi}^{\text{II}(\nu)}_{E,p}, \tilde{\Phi}^{\text{II}(\nu)}_{E',p'})_{\text{II}}&=&(\tilde{\Phi}^{\text{II}(-\nu)}_{E,p}, \tilde{\Phi}^{\text{II}(-\nu)}_{E',p'})_{\text{II}}=8\pi^3a\frac{e^{-\beta p/2}}{\sinh \beta p/2} \delta(E-E')\delta(p-p'), \label{table4}\\
(\tilde{\Phi}^{\text{II}(\nu)}_{E,p}, \tilde{\Phi}^{\text{II}(-\nu)}_{E',p'})_{\text{II}}&=&8\pi^3a\frac{e^{-\beta p/2}}{\sinh \beta p/2} e^{-\pi i \nu} D^{\ast}(E,p)\delta(E-E')\delta(p-p'),  \\
(\tilde{\Phi}^{\text{II}(-\nu)}_{E,p}, \tilde{\Phi}^{\text{II}(\nu)}_{E',p'})_{\text{II}}&=&8\pi^3a\frac{e^{-\beta p/2}}{\sinh \beta p/2} e^{\pi i \nu} D(E,p)\delta(E-E')\delta(p-p'), 
\end{eqnarray}
Here $-\infty <E, E' , p,p' <\infty$. On the other hand we have
\begin{eqnarray}
(\Phi^{\text{II}(\nu)}_{E,p}, \tilde{\Phi}^{\text{II}(\nu)}_{E',p'})_{\text{II}}&=& (\Phi^{\text{II}(-\nu)}_{E,p}, \tilde{\Phi}^{\text{II}(-\nu)}_{E',p'})_{\text{II}}=   0,\\
(\Phi^{\text{II}(\nu)}_{E,p}, \tilde{\Phi}^{\text{II}(-\nu)}_{E',p'})_{\text{II}}&=& (\Phi^{\text{II}(-\nu)}_{E,p}, \tilde{\Phi}^{\text{II}(\nu)}_{E',p'}) _{\text{II}} =0. \label{table8}
\end{eqnarray}
Here $D(E,p)$ is defined by 
\begin{multline}
D(E,p) =\frac{\Gamma( \frac{1+\nu}{2}+i\frac{E+p}{2\sqrt{a}}) \Gamma(\frac{1+\nu}{2}-i\frac{E-p}{2\sqrt{a}}) \Gamma( \frac{1-\nu}{2}+i\frac{E-p}{2\sqrt{a}}) \Gamma(\frac{1-\nu}{2}-i\frac{E+p}{2\sqrt{a}})}
{ \Gamma( \frac{1+\nu}{2}+i\frac{E-p}{2\sqrt{a}}) \Gamma(\frac{1+\nu}{2}-i\frac{E+p}{2\sqrt{a}}) \Gamma( \frac{1-\nu}{2}+i\frac{E+p}{2\sqrt{a}}) \Gamma(\frac{1-\nu}{2}-i\frac{E-p}{2\sqrt{a}}) }
\\
=\frac{\cosh (\beta E/2)+\cos \pi \nu \cosh (\beta p/2)+i\sin \pi \nu \sinh( \beta p/2)}
{\cosh (\beta E/2)+\cos \pi \nu \cosh (\beta p/2)-i\sin \pi \nu \sinh (\beta p/2)}. \label{DEp}
\end{multline}
It can be checked that $D(E,-p)=D(E,p)^{\ast}$ and $D(-E,p)=D(E,p)$.

\section{Near-Horizon Behavior of Normal Modes }
\hspace*{5mm}

The near horizon behavior of the normal mode $\Phi^{\text{II}(\nu)}_{E,p}$ in region II, (\ref{f}), are obtained as follows. For $\eta \sim \frac{2}{\sqrt{a}}$, this mode asymptotes to 
\begin{multline}
\Phi^{\text{II}(\nu)}_{E,p}(\eta,t,x)  
\sim e^{-iEt+ipx}\int_{-\infty}^{\infty}d\zeta\int_0^{\infty}d\mu e^{-i\frac{E}{\sqrt{a}}\zeta-i\frac{p}{\sqrt{a}}\ln \mu} 
J_{\nu}(\mu )\exp\Big\{ i\frac{\sqrt{a}\eta-2}{2} \mu \cosh \zeta \Big\}.   \label{fh}
\end{multline}
By using formula (\ref{Kintegral})
the integration over $\zeta$ in (\ref{fh}) is carried out.
\begin{equation}
\Phi^{\text{II}(\nu)}_{E,p}(\eta,t,x)  \sim 2 e^{-iEt+ipx} \int_0^{\infty} d\mu \mu^{-ip/\sqrt{a}} J_{\nu}(\mu)K_{iE/\sqrt{a}}(-i(\sqrt{a}\eta-2)\mu/2)  \label{JKC2}
\end{equation}
The near horizon behavior is then estimated by using 
\begin{eqnarray}
K_{\nu}(z)&=& \frac{\pi}{2\sin \nu \pi}[I_{-\nu}(z)-I_{\nu}(z)] \nonumber \\
& \stackrel{z\rightarrow 0}{\sim} & \frac{\pi}{2\sin \nu \pi}\Big[ \big(\frac{z}{2}\big)^{-\nu} \frac{1}{\Gamma(1-\nu)}
-\big(\frac{z}{2}\big)^{\nu} \frac{1}{\Gamma(1+\nu)}\Big]  \label{KII}
\end{eqnarray}
as 
\begin{multline}
\Phi^{\text{II}(\nu)}_{E,p}(\eta,t,x)  
\sim e^{-iEt+ipx} e^{-\beta E/4}\Gamma(i\frac{E}{\sqrt{a}}) (\frac{\sqrt{a}\eta-2}{4})^{-iE/\sqrt{a}}\int_0^{\infty} \mu^{-i(E+p)/\sqrt{a}} J_{\nu}(\mu)d\mu \\
+e^{-iEt+ipx} e^{\beta E/4}\Gamma(-i\frac{E}{\sqrt{a}}) (\frac{\sqrt{a}\eta-2}{4})^{iE/\sqrt{a}}\int_0^{\infty} \mu^{i(E-p)/\sqrt{a}} J_{\nu}(\mu)d\mu.
\end{multline}
Finally by using an integration formula
\begin{equation}
\int_0^{\infty}x^{\mu-1}J_{\nu}(ax)dx= 2^{\mu-1}a^{-\mu}\frac{\Gamma((\mu+\nu)/2)}{\Gamma((\nu-\mu)/2+1)}
\end{equation}
the following behavior is obtained.
\begin{multline}
\Phi^{\text{II}(\nu)}_{E,p}(\eta,t,x)  
\sim e^{-iEt+ipx} e^{-\beta E/4} \Gamma(i\frac{E}{\sqrt{a}}) \Big(\frac{\sqrt{a}\eta-2}{4}\Big)^{-iE/\sqrt{a}}
2^{-i(E+p)/\sqrt{a}} \frac{\Gamma(\frac{1}{2}(1+\nu-i\frac{E+p}{\sqrt{a}}))  }{ \Gamma(\frac{1}{2}(1+\nu+i\frac{E+p}{\sqrt{a}}))  }   \\
+e^{-iEt+ipx} e^{\beta E/4} \Gamma(-i\frac{E}{\sqrt{a}}) \Big(\frac{\sqrt{a}\eta-2}{4}\Big)^{iE/\sqrt{a}}
2^{i(E-p)/\sqrt{a}} \frac{\Gamma(\frac{1}{2}(1+\nu+i\frac{E-p}{\sqrt{a}}))  }{ \Gamma(\frac{1}{2}(1+\nu-i\frac{E-p}{\sqrt{a}}))  }  \label{nearhorII}
\end{multline}
This result can also be obtained by using (\ref{Grad}) in (\ref{JKC2}).

Another normal mode, $\tilde{\Phi}^{\text{II}(\nu)}_{E,p}$ defined by (\ref{ftilde}) is related to (\ref{f}) by a relation (\ref{relII}), and the corresponding formulae near the horizon can be obtained by using these relations.

The near horizon behavior of the normal mode in region I, (\ref{phiIEp}), can be obtained in a similar way by using (\ref{KII}) and 
$K_{\nu}(z)= \frac{1}{2}e^{\nu \pi i/2}\int_{-\infty}^{\infty}dt \, e^{-\nu t+iz \sinh t}$: 

\begin{multline}
\Phi^{\text{I}}_{E,p}(t,y,x) \sim  e^{-iEt+ipx}e^{\frac{1}{4}\beta E}2^{-i\frac{p}{\sqrt{a}}} \\
\times \Big[ \Gamma\big(\frac{iE}{\sqrt{a}}\big)\frac{\Gamma(\frac{1+\nu}{2}-i\frac{E+p}{2\sqrt{a}})}{ \Gamma(\frac{1+\nu}{2}+i\frac{E+p}{2\sqrt{a}}) }\big\{\frac{1}{2}(2-\sqrt{a}y)\big\}^{-i\frac{E}{\sqrt{a}}} \\
+\Gamma\big(\frac{-iE}{\sqrt{a}}\big)\frac{\Gamma(\frac{1+\nu}{2}+i\frac{E-p}{2\sqrt{a}})}{ \Gamma(\frac{1+\nu}{2}-i\frac{E-p}{2\sqrt{a}}) }\big\{\frac{1}{2}(2-\sqrt{a}y)\big\}^{i\frac{E}{\sqrt{a}}}\Big].
\end{multline}
The normal mode in region III, (\ref{phiIIIEp}), is also obtained. 
\begin{multline}
\Phi^{\text{III}}_{E,p}(t,y,x) \sim  e^{iEt+ipx}e^{-\frac{1}{4}\beta E}2^{-i\frac{p}{\sqrt{a}}} \\
\times \Big[ \Gamma\big(\frac{iE}{\sqrt{a}}\big)\frac{\Gamma(\frac{1+\nu}{2}-i\frac{E+p}{2\sqrt{a}})}{ \Gamma(\frac{1+\nu}{2}+i\frac{E+p}{2\sqrt{a}}) }\big\{\frac{1}{2}(\sqrt{a}y-2)\big\}^{-i\frac{E}{\sqrt{a}}} \\
+\Gamma\big(\frac{-iE}{\sqrt{a}}\big)\frac{\Gamma(\frac{1+\nu}{2}+i\frac{E-p}{2\sqrt{a}})}{ \Gamma(\frac{1+\nu}{2}-i\frac{E-p}{2\sqrt{a}}) }\big\{\frac{1}{2}(\sqrt{a}y-2)\big\}^{i\frac{E}{\sqrt{a}}}\Big].
\end{multline}

In what follows some inner products of the four mode functions near the horizon are presented. Inner products evaluated at the horizon will be denoted with subscript $H$. 
\begin{eqnarray}
(\Phi^{\text{II}(\nu)}_{E,p}, \Phi^{\text{II}(\nu)}_{E',p'})_{ \text{H} }&=&  (\Phi^{\text{II}(-\nu)}_{E,p}, \Phi^{\text{II}(-\nu)}_{E',p'})_{ \text{H} }   =-16\pi^3 a \delta(E-E')\delta(p-p'), \label{D9}\\
(\tilde{\Phi}^{\text{II}(\nu)}_{E,p}, \tilde{\Phi}^{\text{II}(\nu)}_{E',p'})_{ \text{H} }&=&  (\tilde{\Phi}^{\text{II}(-\nu)}_{E,p}, \tilde{\Phi}^{\text{II}(-\nu)}_{E',p'})_{ \text{H} }   =-16\pi^3 a \delta(E-E')\delta(p-p'), \\
 (\Phi^{\text{II}(\nu)}_{E,p}, \Phi^{\text{II}(-\nu)}_{E',p'})_{ \text{H} }&= &-16\pi^3 a e^{\pi i \nu}
\frac{1+e^{\beta p}+e^{\beta p/2}(e^{\beta E/2}+e^{-\beta E/2})}{e^{\beta p}+e^{2\pi i \nu} +e^{\pi i \nu}e^{\beta p/2}(e^{\beta E/2}+e^{-\beta E/2})} \nonumber \\
&& \cdot \delta(E-E')\delta(p-p'), \\
 (\tilde{\Phi}^{\text{II}(\nu)}_{E,p}, \tilde{\Phi}^{\text{II}(-\nu)}_{E',p'})_{ \text{H} }&= &-16\pi^3 a e^{\pi i \nu}
\frac{1+e^{-\beta p}+e^{-\beta p/2}(e^{\beta E/2}+e^{-\beta E/2})}{e^{-\beta p}+e^{2\pi i \nu} +e^{\pi i \nu}e^{-\beta p/2}(e^{\beta E/2}+e^{-\beta E/2})} \nonumber \\
&& \cdot \delta(E-E')\delta(p-p'), \\
(\Phi^{\text{II}(\nu)}_{E,p}, \tilde{\Phi}^{\text{II}(\nu)}_{E',p'})_{ \text{H} }&=&8\pi^2  \sqrt{a}\delta(E-E')\delta(p-p') E 2^{2ip/\sqrt{a}}|\Gamma(iE/\sqrt{a})|^2(e^{-\beta E/2}-e^{\beta E/2})  \nonumber \\
&&\cdot \frac{  \Gamma(\frac{1}{2}(1+\nu+i\frac{E+p}{\sqrt{a}} )) \Gamma(\frac{1}{2}(1+\nu-i\frac{E-p}{\sqrt{a}} ))}{ \Gamma(\frac{1}{2}(1+\nu-i\frac{E+p}{\sqrt{a}} )) \Gamma(\frac{1}{2}(1+\nu+i\frac{E-p}{\sqrt{a}} )) } \label{D13}
\end{eqnarray}

\section{Solutions to the Matching Conditions}
\hspace*{5mm}
In this Appendix appropriate solutions to the matching conditions (\ref{matching1})-(\ref{matching6}) will be obtained. 
First (\ref{matching1}) and (\ref{matching5}) are analyzed. When $\gamma^{(1)}_4$ is eliminated from these equations, it is found that $\gamma^{(1)}_3$ also disappears, and we have
\begin{equation}
\gamma^{(1)}_2= -\gamma^{(1)}_1 \, \frac{M_1}{M_2}+\frac{b_1}{N^{(1)}_{E,p}M_2} \, C^{-\nu}(-E,p),  \label{gamma12}
\end{equation}
where $M_1$ and $M_2$ are defined in (\ref{M1}) and (\ref{M2}) below. 
This is because the following relations hold
\begin{equation}
C^{-\nu}(-E,p)C^{-\nu}(E,-p)=|\Gamma(i\frac{E}{\sqrt{a}})|^2=C^{-\nu}(-E,-p)C^{-\nu}(E,p).
\end{equation}
Similarly, when $\gamma^{(1)}_2$ is eliminated from  (\ref{matching1}) and (\ref{matching3}), then $\gamma^{(1)}_1$ also dropps out. 
\begin{equation}
\gamma^{(1)}_4= \gamma^{(1)}_3 \, \frac{M_3}{M_2}-\frac{b_1}{N^{(1)}_{E,p}M_2} \, C^{\nu}(-E,p),   \label{gamma14}
\end{equation}
where $M_3$ is defined in (\ref{M3}). Here 
\begin{eqnarray}
M_1 & \equiv& C^{-\nu}(-E,p)C^{\nu}(E,-p)-C^{-\nu}(E,p)C^{\nu}(-E,-p), \label{M1}\\
M_2 &\equiv& C^{-\nu}(-E,p)C^{\nu}(E,p)-C^{-\nu}(E,p)C^{\nu}(-E,p), \label{M2} \\
M_3 & \equiv & C^{\nu}(-E,p)C^{-\nu}(E,-p)-C^{\nu}(E,p)C^{-\nu}(-E,-p). \label{M3}
\end{eqnarray}
$b_1$ is defined by removing $e^{-ipx}V^{-iE/\sqrt{a}}$ from the right-hand side of (\ref{matching1})
\begin{equation}
b_1 \equiv \sqrt{1-e^{-\beta E}}2^{i(p-E)/\sqrt{a}}\frac{\Gamma(\frac{1+\nu}{2}-i\frac{E-p}{2\sqrt{a}})}{ \Gamma(\frac{1+\nu}{2}+i\frac{E-p}{2\sqrt{a}}) } e^{\beta E/4}
\Gamma(i\frac{E}{\sqrt{a}})    \label{b1}
\end{equation}

In order to make the norm (\ref{norm1}) with $i=1$ take the form (\ref{correctnorm}), it is assumed that $\gamma^{(1)}_1$ and $\gamma^{(1)}_3$ satisfy the following equations,
\begin{eqnarray}
& |\gamma^{(1)}_1+\gamma^{(1)}_3 e^{-\pi i \nu}D^{\ast}(E,p)|^2 = F^{(1)}_1(E,p), \label{assump1}\\
& |\gamma^{(1)}_1 +\gamma^{(1)}_3 e^{-\pi i \nu}+J^{(1)}|^2= F^{(1)}_2(E,p) \label{assump2}
\end{eqnarray}
where $F^{(1)}_{1,2}(E,p)$ are functions which satisfy for $E, p \geq 0$
\begin{itemize}
\item $F_1^{(1)}(E,p) \geq 0$ and $F_2^{(1)}(E,p) \geq 0$ 
\item $e^{-\beta p/2}F_1^{(1)}(E,p) - e^{\beta p/2} F_2^{(1)}(E,p) =e^{\beta p/2}-e^{-\beta p/2}$, 
\end{itemize}
where
\begin{equation}
J^{(1)} \equiv  \frac{b_1}{N^{(1)}_{E,p}}\frac{Ee^{-\beta E/2}e^{-\pi i \nu}}{2\pi \sqrt{a}}C^{\nu}(-E,p)[e^{\beta (E-p)/2}+e^{\pi i \nu}]
\end{equation}
Then it can be shown that (\ref{correctnorm}) with $i=1$ holds. 
First, note that (\ref{assump1}) and (\ref{assump2}) can be rewritten as 
\begin{eqnarray}
& \gamma^{(1)}_1+\gamma^{(1)}_3 e^{-\pi i \nu}D^{\ast}(E,p)=(F^{(1)}_1)^{1/2}e^{i\delta_1}, \label{g13}\\
& \gamma^{(1)}_1 +\gamma^{(1)}_3 e^{-\pi i \nu} =( F^{(1)}_2)^{1/2} e^{i\delta_2}-J^{(1)} \label{g13s}
\end{eqnarray}
Here $\delta_{1,2}(E,p)$ are arbitrary real constants. These equations are solved for $\gamma^{(1)}_{1,3}$ as
\begin{eqnarray}
\gamma^{(1)}_1 &=& K \, e^{-\pi i\nu} [(F^{(1)}_1)^{1/2} e^{i\delta_1}-D^{\ast}(E,p)((F^{(1)}_2)^{1/2}e^{i\delta_2}-J^{(1)})], \\
\gamma^{(1)}_3 &=& K \, [-(F^{(1)}_1)^{1/2} e^{i\delta_1}+(F^{(1)}_{2})^{1/2}e^{i\delta_2}-J^{(1)}], 
\end{eqnarray}
where 
\begin{equation}
K \equiv  \frac{[e^{\beta (E-p)/2}+e^{\pi i\nu}][e^{\beta(E+p)/2}+e^{-\pi i \nu}]}
{4i e^{\beta E/2} \, \sinh(\beta p/2) e^{-\pi i \nu} \, \sin \pi \nu}.
\end{equation}
Now by using (\ref{gamma12}) and (\ref{gamma14}) $\gamma^{(1)}_2$ and $\gamma^{(1)}_4$ are also evaluated explicitly.  The results are given in (\ref{gamma2}) and (\ref{gamma4}). 
It is now straightforward to show that 
\begin{eqnarray}
(\psi^{(1)}_{E,-p},\psi^{(1)}_{E',-p'})&=(\psi^{(3)}_{E,-p},\psi^{(3)}_{E',-p'})= |N^{(1,3)}_{E,p}|^2(e^{\beta p/2}-e^{-\beta p/2})^{-1}H(E,p) \nonumber\\
 & \qquad \times 16\pi^3 a \delta(E-E')\delta(p-p'),
\end{eqnarray}
where 
\begin{equation}
H(E,p)= e^{-\beta p/2}F^{(1)}_1(E,p)-e^{\beta p/2}F^{(1)}_2(E,p)=e^{\beta p/2}-e^{-\beta p/2},
\end{equation}
by using (\ref{table1})-(\ref{table8}). Then it is necessary to set $|N^{(1)}_{E,p}|=  |N^{(3)}_{E,p}|=1$ in order to enforce a correct normalization of the norms. 

Also by comparing (\ref{matching1}) and (\ref{matching3}) it is found that 
\begin{equation}
\frac{b_1}{N^{(1)}_{E,p}}=\frac{b_3}{N^{(3)}_{E,p}}, \label{N13}
\end{equation}
where
 $b_3$ is defined by removing $e^{ipx}U^{-iE/\sqrt{a}}$ from the right-hand side of (\ref{matching3})
\begin{equation}
b_3 \equiv \sqrt{1-e^{-\beta E}}2^{-i(p+E)/\sqrt{a}}\frac{\Gamma(\frac{1+\nu}{2}-i\frac{E+p}{2\sqrt{a}})}{ \Gamma(\frac{1+\nu}{2}+i\frac{E+p}{2\sqrt{a}}) } e^{\beta E/4}
\Gamma(i\frac{E}{\sqrt{a}}).    \label{b3}
\end{equation}
Then (\ref{b1}) and (\ref{N13}) imply that $N^{(1,3)}_{E,p}$ must have the forms
\begin{eqnarray}
N^{(1)}_{E,p} &=& 2^{i(p-E)/\sqrt{a}}\frac{\Gamma(\frac{1+\nu}{2}-i\frac{E-p}{2\sqrt{a}})}{ \Gamma(\frac{1+\nu}{2}+i\frac{E-p}{2\sqrt{a}}) } \, e^{i\delta_N}, \label{N1}\\
N^{(3)}_{E,p} &=& 2^{-i(p+E)/\sqrt{a}}\frac{\Gamma(\frac{1+\nu}{2}-i\frac{E+p}{2\sqrt{a}})}{ \Gamma(\frac{1+\nu}{2}+i\frac{E+p}{2\sqrt{a}}) } \, e^{i\delta_N}
\end{eqnarray}
where $\delta_N(E,p)$ is an arbitrary real number. 

Exactly the same way, solution to the matching conditions for $\psi^{(2)}_{E,p}$ and $\psi^{(4)}_{E,p}$ are solved and $\gamma^{(2)}_n$ $(n=1,2,3,4)$ are obtained. By using (\ref{matching2}) and (\ref{matching6}) $\gamma^{(2)}_2$ and $\gamma^{(2)}_4$ are expressed in terms of $\gamma^{(2)}_1$ and $\gamma^{(2)}_3$ as
\begin{eqnarray}
\gamma^{(2)}_2 &=& -\gamma^{(2)}_1\frac{M_1}{M_2}+\frac{b_2}{N^{(2)}_{E,p}}\frac{C^{-\nu}(-E,p)}{M_2}, \label{gamma22}\\
\gamma^{(2)}_4 &=& \gamma^{(2)}_3\frac{M_3}{M_2}-\frac{b_2}{N^{(2)}_{E,p}}\frac{C^{\nu}(-E,p)}{M_2} \label{gamma24}
\end{eqnarray}
In order to make the inner product of $\psi^{(2)}_{E,p}$ positive definite and normalized to unity
the following two conditions are imposed
\begin{eqnarray}
& |\gamma^{(2)}_1+\gamma^{(2)}_3 e^{-\pi i \nu}D^{\ast}(E,p)|^2=F^{(2)}_1(E,p), \\
& |\gamma^{(2)}_1+\gamma^{(2)}_3 e^{-\pi i\nu}+J^{(2)}|^2= F^{(2)}_2(E,p)
\end{eqnarray}
Here $F^{(2)}_{1,2}(E,p)$ satisfty the same conditions as those for $F^{(1)}_{1,2}$. 
These equations are solved as
\begin{eqnarray}
\gamma^{(2)}_1 &=& K e^{-\pi i\nu} [ (F^{(2)}_1)^{1/2} e^{i\delta_3}-D^{\ast}(E,p)((F^{(2)}_2)^{1/2}e^{i\delta_4}-J^{(2)})], \\
\gamma^{(2)}_3 &=&K [-(F^{(2)}_1)^{1/2}e^{i\delta_3}+(F^{(2)}_2)^{1/2}e^{i\delta_4}-J^{(2)}]
\end{eqnarray}
Here $e^{i\delta_3(E,p)}$ and $e^{i\delta_4(E,p)}$ are phase factors and 
\begin{equation}
J^{(2)} = \frac{b_2}{N^{(2)}_{E,p}} \frac{Ee^{-\beta E/2}e^{-\pi i \nu}}{2\pi \sqrt{a}}C^{\nu}(-E,p)[e^{\beta (E-p)/2}+e^{\pi i \nu}]
\end{equation}
Then (\ref{gamma22}) and (\ref{gamma24}) determine $\gamma^{(2)}_2$ and $\gamma^{(2)}_4$, and it can be shown that $(\psi^{(2)}_{E,p},\psi^{(2)}_{E',p'})=(\psi^{(4)}_{-E,-p},\psi^{(4)}_{-E',-p'})= 16\pi^3a |N^{(2,4)}_{E,p}|^2\delta(E-E')\delta(p-p')$ by using (\ref{table1})-(\ref{table8}). 
 $N^{(2,4)}_{E,p}$ are phase factors which satisfy $b_2/N^{(2)}_{E,p}=b_4/N^{(4)}_{E,p}$. Since $b_4=b_1$ and $b_3=b_2$, it follows that  $N^{(2)}_{E,p}/N^{(4)}_{E,p}=b_2/b_4=b_3/b_1=N^{(3)}_{E,p}/N^{(1)}_{E,p}$, and we obtain $N^{(2)}_{E,p}=N^{(3)}_{E,p}e^{i(\delta'_N-\delta_N)}$ and $N^{(4)}_{E,p}=N^{(1)}_{E,p}e^{i(\delta'_N-\delta_N)}$, where $\delta'_N(E,p)$ is a real constant.


\end{document}